\documentclass[preprint,superscriptaddress,preprintnumbers,floatfix]{revtex4-1}

\usepackage{amsthm}
\usepackage{amsmath}
\usepackage{mathrsfs}
\usepackage{amssymb}
\usepackage{wasysym}
\usepackage{graphicx}
\usepackage{varwidth}
\usepackage{url}
\usepackage{dsfont}
\usepackage{float}
\usepackage{bm}
\usepackage{ifthen}
\usepackage[usenames,dvipsnames]{color}
\usepackage{mathrsfs}
\usepackage{hyperref}
\usepackage{float}
\usepackage{afterpage}
\usepackage[caption=false]{subfig}
\usepackage{textcomp}

\newsavebox{\twosubbox}
\usepackage{physics}
\usepackage{xcolor}
\usepackage{fancyhdr}
\usepackage{ulem}

\usepackage{enumitem}
\usepackage{mathtools}
\usepackage{algorithm,algpseudocode}
\algnewcommand{\LeftComment}[1]{\Statex \hspace{3em}\(\triangleright\) #1}
\usepackage{multirow}

\usepackage{amsthm}

\theoremstyle{definition}

\begin{document}
\preprint{CaltechAUTHORS:0b9qh-jt654; FERMILAB-PUB-24-0184-ETD}
\title{Long-range wormhole teleportation}
\begin{abstract}
We extend the protocol of \citet{gao2021traversable}
to allow wormhole teleportation between two entangled copies of the
Sachdev-Ye-Kitaev (SYK) model~\cite{PhysRevLett.70.3339,kitaev2015simple}
communicating only through a classical channel. We demonstrate in finite $N$ simulations that the protocol exhibits the characteristic holographic features of
wormhole teleportation discussed and summarized in \citet{jafferis2022traversable}. We review and exhibit in detail how these holographic
features relate to size winding which, as first shown by \citet{brown2021quantum}, \citet{nezami2021quantum},
encodes a dual description of wormhole teleportation.
\end{abstract}

\author{Joseph D. Lykken}
\affiliation{Quantum Division, Fermi National Accelerator Laboratory, Batavia, IL, USA}

\author{Daniel Jafferis}
\affiliation{Center for the Fundamental Laws of Nature, Harvard University, Cambridge, MA, USA}

\author{Alexander Zlokapa}
\affiliation{Center for Theoretical Physics, Massachusetts Institute of Technology, Cambridge, MA, USA}

\author{David K. Kolchmeyer}
\affiliation{Center for Theoretical Physics, Massachusetts Institute of Technology, Cambridge, MA, USA}

\author{Samantha I. Davis}
\affiliation{Division of Physics, Mathematics and Astronomy, Caltech, Pasadena, CA, USA}
\affiliation{Alliance for Quantum Technologies (AQT), California Institute of Technology, Pasadena, CA, USA}

\author{Hartmut Neven}
\affiliation{Google Quantum AI, Venice, CA, USA} 

\author{Maria Spiropulu}
\affiliation{Division of Physics, Mathematics and Astronomy, Caltech, Pasadena, CA, USA}
\affiliation{Alliance for Quantum Technologies (AQT), California Institute of Technology, Pasadena, CA, USA}


\collaboration{QCCFP Collaboration}
\noaffiliation

\maketitle

\section{Introduction}\label{sec:intro}

Building on the work of \citet{gao2017traversable}, \citet{maldacena2017diving}, and
others \cite{Maldacena:2001kr,Hayden:2007cs,Sachdev:2010um,Shenker:2013pqa,
Shenker:2013yza,Maldacena:2016hyu,Susskind:2016jjb,Susskind:2017nto,Susskind:2017ney},
the wormhole teleportation protocol developed by 
\citet{gao2021traversable} is a concrete realization of the 
$ER$$=$$EPR$ hypothesis of \citet{coolhorizons}. The protocol consists of a series of quantum gate operations, of the type
routinely performed now in the laboratory on various kinds of quantum processors. Importantly there is a well-defined semi-classical limit in which this teleportation protocol has an equivalent holographic description as coherent transmission of quantum states
through a traversable wormhole. 

The basic semi-classical picture
resembles an Einstein-Rosen bridge connecting two black holes.
In the traversable wormhole teleporation  protocol \cite{gao2021traversable} the role of the two black holes is played by two entangled copies of the Sachdev-Ye-Kitaev (SYK) model~\cite{PhysRevLett.70.3339,kitaev2015simple}.
The SYK model is a quantum mechanical system of $N$
Majorana fermions interacting $q$ at a time in all possible combinations
with Hamiltonian:
\begin{align}
\label{eq:sykH}
    H_{L,R} = \sum_{1 \leq j_1 < \dots < j_q \leq N} J_{j_1\dots j_q} \psi^{\ell,r}_{j_1}\dots \psi^{\ell, r}_{j_q},
\end{align}
where we use the Left-Right symbols $L,R$ to distinguish the two SYK copies
with Majorana fermions corresponding labeled by $\ell,r$.
The couplings $J_{j_1\dots j_q}$ are the same for both copies, and are
drawn randomly from a Gaussian distribution with mean zero and variance 
\begin{align}
\label{eq:variance}
\langle \left( J_{j_1\dots j_q} \right)^2 \rangle
= \frac{2^{q-1}(q-1)!}{qN^{q-1}} {\cal J}^2
\end{align}
where ${\cal J}$ is a coupling constant with dimensions
of energy.
The SYK Hamiltonian resembles the dynamics of a black hole in that it 
is an efficient scrambler. Scrambling in this context refers to the fact that quantum information injected into the system via a simple operator rapidly delocalizes into
multipartite entanglement. One can track scrambling via operator
growth over time. Like black holes, the SYK model saturates the upper bound on the Lyapunov exponent describing operator growth \cite{maldacena2016bound,Maldacena:2016hyu}, 
in the limit
$N$$\to$$\infty$,  $\beta{\cal J}$$\to$$\infty$, 
$N/\beta{\cal J}$$\to$$\infty$.

The dynamics of the SYK model has no spatial dependence, and  the
action of Majorana fermion operators in zero spatial dimensions is equivalent
to strings of Pauli operators via the Jordan-Wigner transformation. As a result the
quantum dynamics of SYK could be realized on a variety of digital or analog quantum processors
independently of the details of the hardware implementation.
To the extent that one can then invoke a holographic description 
of wormhole teleportation as traversing
a wormhole, the space being traversed is not the space of the laboratory,
but rather an emergent space arising from the dynamics of quantum entanglement. This makes wormhole teleportation a promising window for probing
features of quantum gravity away from the semi-classical regime.

It was demonstrated in \cite{gao2021traversable} that, in a semi-classical limit with
$N$$\to$$\infty$, $q$$\to$$\infty$, $\beta{\cal J}$$\to$$\infty$,
traversable wormhole teleportation can have perfect fidelity, and exhibits a number of
qualitative features related to the holographic description with a wormhole.
The most obvious of these is that, unlike the 
standard Alice-Bob teleportation protocol \cite{Bennett:1992tv}
employed in quantum networks (see, e.g. \cite{Valivarthi:2020yax}) 
wormhole teleportation completes after a 
dynamically determined time interval, equivalent in the dual description to
the time taken, as measured by an external observer, to transit the wormhole.
In \cite{jafferis2022traversable} it was shown, both explicitly on a quantum processor and in more detail in a simulation on a classical computer, that several qualitative features of wormhole teleportation persist well away from the semi-classical regime (e.g. $N$$=$$10$, $q$$=$$4$, $\beta{\cal J}$$=$$4$), at the expense of reduced fidelity. 

An additional
complication noted in \cite{gao2021traversable} is that when the holographic
teleportation has less than perfect fidelity, it competes with
other non-holographic effects. The first  is a ``direct swapping'' of quantum information intrinsic to the protocol, which dominates at early times and degrades rapidly due to the same scrambling dynamics that induce wormhole teleportation at its characteristic time scale. The second is a many-body effect, dubbed ``peaked-size teleportation'' in \cite{schuster2021manybody}, which always dominates at late times after the characteristic completion time of the wormhole teleportation, as well as at infinite temperature.
An additional possible mechanism, not seen in the examples discussed in this paper,
is ``teleportation through thermalization", a many-body
mechanism discussed recently in \cite{Gao:2023gta}.

The wormhole teleportation protocol of
\citet{gao2021traversable} includes applying, over some time interval, an explicit 
unitary operator exp$(i\mu V)$ connecting the two entangled SYK systems. Here
$\mu$ is a coupling constant and $V$ is proportional to
a sum of Majorana fermion bilinears $\psi^\ell_j\psi^r_j$.
In the holographic description
this interaction introduces a negative energy pulse 
resulting in a Shapiro advance that makes the wormhole traversable; this in turn implies the requirement that the coupling
constant $\mu$ must be negative. As already observed in
\cite{gao2017traversable}, \cite{maldacena2017diving}, and 
\cite{gao2021traversable}, traversability of the wormhole does not 
uniquely determine the form of the Left-Right interaction.

The wormhole teleportation described so far has a key difference from
the standard Alice-Bob teleportation that prevents it from operating
between separated systems. Standard Alice-Bob
teleportation is based on an entangled Bell pair shared
between Alice and Bob. In that protocol Alice performs a Bell state measurement
jointly on her qubit and a second message qubit; she then calls Bob via a classical
channel and instructs him to perform particular gate operations on his qubit,
the choice of which depends on Alice's measurement outcome. The teleportation of
the message qubit  completes successfully, without any quantum interaction
between Alice and Bob. The wormhole teleportation protocol described in
\cite{gao2021traversable} and performed on a Sycamore quantum processor \cite{jafferis2022traversable} has no classical channel; it
resembles instead a variation of the standard Alice-Bob protocol 
(see, e.g., \cite{Lykken:2020xtx}) where Alice's
measurement and the classical channel to Bob is replaced by a quantum channel consisting of two control gates (CNOT and CZ) that directly connect Alice and Bob. The quantum channel
consists entirely of control gates with Alice's qubits as the controls, hence the
deferred measurement principle \cite{Nielsen:2012yss} guarantees that the two protocols achieve
identical results.

The wormhole teleportation protocol in \cite{gao2021traversable} does not have
an equivalent variant using a classical channel because of
the particular form chosen for the Left-Right coupling $V$. Thus to achieve long-range wormhole teleportation we need to first modify
$V$. We  use a modified Left-Right coupling suggested by 
\citet{nezami2021quantum} (see also \cite{maldacena2017diving} and
\cite{Kourkoulou:2017zaj}) and
demonstrate that we can still achieve wormhole teleportation with the previously identified holographic features. 
As a final step we modify the protocol to make use of a classical channel. 

The outline of this paper is as follows. In section \ref{sec:notation} we
establish notation and review the basics of the wormhole teleportation
protocol of \cite{gao2021traversable}.
In section \ref{sec:warmup}
we warm up with the case of the Gao-Jafferis protocol at
infinite temperature and no time allowed for scrambling; here the final state can be computed analytically. We observe
no wormhole teleportation but instead the Left-Right interaction induces
a Left-Right swap, leading to nonzero quantum information transfer. This
is sufficient to see explicitly that the protocol of \cite{gao2021traversable}
cannot be replaced with a long-range classical channel. 
In section \ref{sec:lrprotocol} we replace the Left-Right interaction operator $V$ with a different operator $V^b$, and show that the basic features of traversable wormhole teleportation also appear in this modified protocol.
We show that the Left-Right swap is absent for this
long-range protocol.
In section \ref{sec:sizewinding} we review
the size winding description of traversable wormhole
teleportation, and demonstrate how it works using either the
operator $V$ or the operator $V^b$ to define ``size". We show
that both cases exhibit an approximate $SL(2,R)$ symmetry,
which is closely related to size winding in the holographic 
description \cite{Maldacena:2018lmt,brown2021quantum,nezami2021quantum}.
From the size winding analysis we are able to extract finite-$N$
Lyapunov exponents and compare them to the semi-classical limit.
In section \ref{sec:timeordering}
we discuss the holographic feature of
causal time ordering, and show that it occurs in the 
long-range teleportation protocol.
In section \ref{sec:classical} we demonstrate how to further
modify the protocol to include a classical channel. This involves
measurement of Alice's qubits on the left, and displays some interesting
features of multipartite entanglement. 
Finally, in section \ref{sec:outlook} we discuss the
possibility and challenges of implementing long-range
wormhole teleportation on quantum hardware, 
perhaps over quantum networks currently under
development. 
We provide in Appendix \ref{sec:commuting}
a comparison between the SYK-based protocols discussed here
and the commuting models recently studied by \citet{Gao:2023gta}.

\section{Notation and definitions}\label{sec:notation}

We choose units such that $\cal{J}$$=$$\hbar$$=$$c$$=$1, thus when we write time $t$ as a dimensionless number we actually mean $t\cal{J}/\hbar$, and when we write the inverse temperature $\beta$ as a dimensionless number we actually mean $\beta$$\cal{J}$.

We mostly use the notation and conventions of \citet{gao2021traversable}.
The two entangled SYK systems each have $N$ Majorana fermions;
we label Alice's by $\psi^\ell_i$ and Bob's by 
$\psi^r_i$, $i=0,\ldots N$$-$$1$. The Majorana operator normalization is
implied by the anti-commutation
relations:
\begin{align}
\label{eq:commutators}
\{ \psi^\ell_i, \psi^\ell_j \}  =
\{ \psi^r_i, \psi^r_j \} = \delta_{ij}
\end{align}
Expectation values at $\beta$$=$$0$ are taken with respect
to the maximally entangled state $\ket{I}$ that satisfies
\begin{align}
\label{eq:Istate}
\psi^\ell_j \ket{I} = -i\psi^r_j \ket{I} \; ;\quad
\psi^r_j \ket{I} = i\psi^\ell_j \ket{I} 
\end{align}
for all of the left and right Majoranas $\psi^\ell_j$,
$\psi^r_j$, $j = 0,\ldots ,N$$-$$1$. 
At finite $\beta$ and $t$$=$$0$ we will
take expectation values with respect to the normalized
thermofield double state $\ket{\rm tfd}$ defined by
\begin{align}
\label{eq:tfd}
\ket{\rm tfd} \equiv \rho_\beta^{1/2} \ket{I}
= 
Z^{-1/2}_\beta{\rm e}^{-\beta H_L/2} \ket{I} = 
Z^{-1/2}_\beta{\rm e}^{-\beta H_R/2} \ket{I}
\end{align}
where $\beta$ is the inverse temperature, $Z_\beta$ is the corresponding partition function, and we have used the fact that 
$(H_L - H_R)\ket{I} = 0$.

The wormhole teleportation protocol of
\citet{gao2021traversable} introduced an explicit coupling
between the left and right Majoranas 
exp$(i\mu V)$, where the Hermitian operator $V$ is defined by
\begin{align}
\label{eq:Vdef}
V  = 
i\sum_{j=0}^{N-1} \, 
\psi^\ell_j \psi^r_j 
\end{align}
and $\mu$ is a coupling constant. In the simplest version of the
protocol this interaction is applied at a single time slice
$t$$=$$0$, but we will also consider, following \cite{jafferis2022traversable},
extending the interaction over multiple time steps.
Note that in 
\cite{gao2021traversable} $V$ was defined with an additional factor of
$1/(qN)$. This means that to compare a $\mu$ value used here
to a $\mu$ value in \cite{gao2021traversable}, one should
first multiply it by $qN$. Our convention for $\mu$ matches with
\citet{Gao:2023gta}.

The quantum information to be teleported initially
resides in a Bell state pair of qubits denoted $R$,$Q$, and the
goal is to transfer the quantum information to a readout qubit $T$.
At some ``injection time'' $-t_0$ we want to swap the qubit $Q$ with
one of the qubits implementing the Left SYK dynamics, and at the
``extraction time'' $t_1$ we need to swap out into $T$ from one of the 
qubits that implement the Right SYK dynamics.
The wormhole teleportation protocol uses the following complex fermions for swapping:
\begin{align}
\label{eq:chidef}
\chi_\ell  = \frac{1}{\sqrt{2}}\left( \psi^\ell_0 + i \psi^\ell_1 \right)\; ; \quad
\{ \chi_\ell,\chi_\ell^\dagger \} = 1 \; ; \quad \chi_\ell^2 = (\chi_\ell^\dagger)^2 = 0\\
\chi_r  = \frac{1}{\sqrt{2}}\left( \psi^r_0 + i \psi^r_1 \right) \; ; \quad
\{ \chi_r,\chi_r^\dagger \} = 1 \; ; \quad \chi_r^2 = (\chi_r^\dagger)^2 = 0
\end{align}
As in eqn. 2.6 of \cite{gao2021traversable}, the Left swap operation $S_{Q\ell}$ from the
qubit Q into the Left Majoranas, and the Right swap operation $S_{Tr}$ from
the Right Majoranas into the readout qubit T, are given by:
\begin{align}
\label{eq:swapdef}
S_{Q\ell} = 
\begin{pmatrix}
1 & 0 \\
0 & 0
\end{pmatrix}_Q
\otimes \chi_\ell \chi_\ell^\dagger +
\begin{pmatrix}
0 & 1 \\
0 & 0
\end{pmatrix}_Q
\otimes \chi_\ell^\dagger + 
\begin{pmatrix}
0 & 0 \\
1 & 0
\end{pmatrix}_Q
\otimes \chi_\ell + 
\begin{pmatrix}
0 & 0 \\
0 & 1
\end{pmatrix}_Q
\otimes \chi_\ell^\dagger \chi_\ell \\
S_{Tr} = 
\begin{pmatrix}
1 & 0 \\
0 & 0
\end{pmatrix}_T
\otimes \chi_r \chi_r^\dagger +
\begin{pmatrix}
0 & 1 \\
0 & 0
\end{pmatrix}_T
\otimes \chi_r^\dagger + 
\begin{pmatrix}
0 & 0 \\
1 & 0
\end{pmatrix}_T
\otimes \chi_r + 
\begin{pmatrix}
0 & 0 \\
0 & 1
\end{pmatrix}_T
\otimes \chi_r^\dagger \chi_r
\end{align}
The initial $RQ$ state is taken to be the Bell state $(\ket{00} + \ket{11})/\sqrt{2}$, while the
initial state of the readout qubit $T$ is $\ket{0}$.
The full quantum state at time $t$$=$$0$ is thus the following tensor product state
with $N$$+$$3$ qubits:
\begin{align}
\label{eq:initial}
\ket{t=0} = \frac{1}{\sqrt{2}}(\ket{00} + \ket{11})
\otimes \ket{\rm tfd} \otimes \ket{0}
\end{align}
and the wormhole teleportation protocol produces the following final state:
\begin{align}
\label{eq:final}
\ket{\rm final} =    S_{Tr}\,{\rm e}^{-iH_Rt_1}\, {\rm e}^{i\mu V}\,
{\rm e}^{-iH_Lt_0}\, S_{Q\ell}\,{\rm e}^{iH_Lt_0} \ket{t=0}
\end{align}
Thus, starting at $t$$=$$0$, we time evolve backwards with the Left SYK
hamiltonian to the injection time $-t_0$, swap in the quantum information of qubit
$Q$, time evolve back to $t$$=$$0$, apply the Left-Right interaction,
time evolve with the Right SYK hamiltonian to the extraction time $t_1$,
then swap out into the readout qubit $T$. In the holographic picture the
two copies of the SYK system (Left and Right) time evolve together and, as for the Einstein-Rosen
bridge, forward time evolution with $H_L$ is like backward time evolution with
$H_R$, thus what $H_L$ scrambles, $H_R$ unscrambles. In the protocol we
only include the pieces of the time evolution that affect the teleportation outcome.

As shown in \cite{gao2021traversable}, it
follows that the reduced density matrix of $TR$ in the final state is given by:
\begin{align}
\label{eq:rhoRTdef}
\rho_{TR} = \frac{1}{2}
\begin{pmatrix}
\rho_{11} & 0 & 0 & \rho_{14} \\
0 & \rho_{22} & \rho_{23} & 0 \\
0 & \rho_{23}^* & \rho_{33} & 0 \\
\rho_{14}^* & 0 & 0 & \rho_{44}
\end{pmatrix}
\end{align}
where here we have absorbed the time evolution into
$\chi_\ell$$\to$$\chi_\ell(-t_0)$, $\chi_r$$\to$$\chi_r(t_1)$, and where
\begin{align}
\label{eq:rhodefsa}
\rho_{11} &= \left\langle \chi_\ell\chi_\ell^\dagger 
\,{\rm e}^{-i\mu V} \chi_r\chi_r^\dagger \,{\rm e}^{i\mu V} 
\chi_\ell\chi_\ell^\dagger \right\rangle +
\left\langle \chi_\ell^\dagger 
\,{\rm e}^{-i\mu V} \chi_r\chi_r^\dagger \,{\rm e}^{i\mu V}
\chi_\ell \right\rangle \nonumber\\
\rho_{14} &= \left\langle \chi_\ell 
\,{\rm e}^{-i\mu V} \chi_r^\dagger \,{\rm e}^{i\mu V}
\chi_\ell\chi_\ell^\dagger \right\rangle +
\left\langle \chi_\ell^\dagger\chi_\ell 
\,{\rm e}^{-i\mu V} \chi_r^\dagger \,{\rm e}^{i\mu V}
\chi_\ell \right\rangle \nonumber\\
\rho_{22} &= \left\langle \chi_\ell
\,{\rm e}^{-i\mu V} \chi_r\chi_r^\dagger \,{\rm e}^{i\mu V} 
\chi_\ell^\dagger \right\rangle +
\left\langle \chi_\ell^\dagger\chi_\ell 
\,{\rm e}^{-i\mu V} \chi_r\chi_r^\dagger \,{\rm e}^{i\mu V} 
\chi_\ell^\dagger\chi_\ell \right\rangle \nonumber\\
\rho_{23} &= \left\langle \chi_\ell\chi_\ell^\dagger 
\,{\rm e}^{-i\mu V} \chi_r^\dagger \,{\rm e}^{i\mu V}
\chi_\ell^\dagger \right\rangle +
\left\langle \chi_\ell^\dagger
\,{\rm e}^{-i\mu V} \chi_r^\dagger \,{\rm e}^{i\mu V}
\chi_\ell^\dagger\chi_\ell \right\rangle \\
\rho_{33} &= \left\langle \chi_\ell\chi_\ell^\dagger
\,{\rm e}^{-i\mu V} \chi_r^\dagger\chi_r \,{\rm e}^{i\mu V} 
\chi_\ell\chi_\ell^\dagger \right\rangle +
\left\langle \chi_\ell^\dagger
\,{\rm e}^{-i\mu V} \chi_r^\dagger\chi_r \,{\rm e}^{i\mu V} 
\chi_\ell \right\rangle \nonumber\\
\rho_{44} &= \left\langle \chi_\ell
\,{\rm e}^{-i\mu V} \chi_r^\dagger\chi_r \,{\rm e}^{i\mu V} 
\chi_\ell^\dagger \right\rangle +
\left\langle \chi_\ell^\dagger\chi_\ell
\,{\rm e}^{-i\mu V} \chi_r^\dagger\chi_r \,{\rm e}^{i\mu V} 
\chi_\ell^\dagger\chi_\ell \right\rangle \nonumber
\end{align}

\begin{figure}[htb]
\begin{center}
\includegraphics[width=0.32\textwidth]{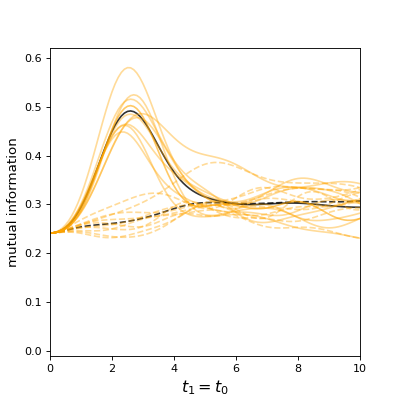}
\includegraphics[width=0.32\textwidth]{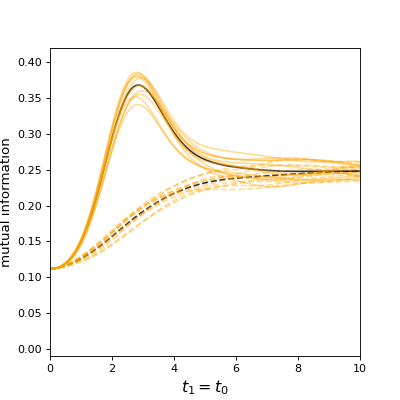}
\includegraphics[width=0.32\textwidth]{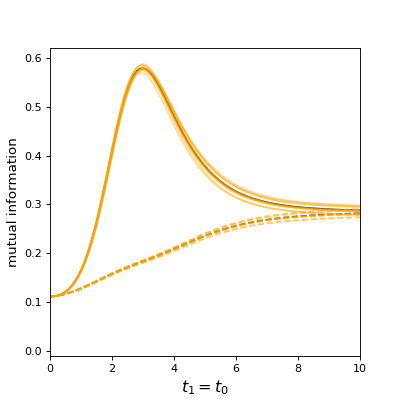}
\protect\caption{{\bf Left:} Mutual information (computed using the second Renyi entropies) 
between the qubits $R$ and $T$ after
running the wormhole teleportation protocol with injection at time
$-t_0$ and extraction at time $t_1$$=$$t_0$. Results from ten
instantiations of the $N$$=$$10$ SYK model with $q$$=$$4$, 
$\cal J$$=$$1$, $\beta$$=$$4$,
using the Left-Right interaction
$V$ with coupling $\mu $$=$$-0.3$ (solid)/$\mu $$=$$+0.3$ (dashed).
The orange lines are the individual instantiations, and the
black lines are the mean. {\bf Middle:} The same for $N$$=$$14$,
except $\mu$$=$$\pm0.2$.
{\bf Right:} The same for $N$$=$$24$, except $\mu$$=$$\pm0.2$.
\label{fig:self-average}}
\end{center}
\vspace{-0.1cm}
\end{figure}

Figure \ref{fig:self-average} shows results typical of the
wormhole teleportation protocol at finite $N$. Plotted is the
mutual information $I(R$$:$$T)$
(in units of log2) between the qubits $R$ and $T$ as a function
of time, averaged over 10 instantiations of the SYK model
random couplings, and taking the injection and extraction times to be equal in magnitude: $t_0$$=$$t_1$. The figure shows results for both
negative and positive values of the Left-Right interaction coupling
$\mu$.
The quantum information shared between
$R$ and $T$ has three different contributions: (i) holographic
wormhole teleportation, which produces a peak in the
mutual information at a characteristic time, but only when $\mu$
is chosen negative; (ii) non-holographic direct swapping of quantum information
between the Left and Right systems, described in detail in the next section,  which occurs maximally for $t_0=t_1=0$
and is symmetric under the sign chosen for $\mu$; (iii) 
non-holographic many-body ``peaked-size teleportation'' , discussed in detail in subsection \ref{sec:peaksize}, which
occurs maximally at late times and is symmetric under the sign chosen for $\mu$. 

As seen in Figure \ref{fig:self-average}, for $N$$=$$10$ and 14 there is substantial
variation in the mutual information plot from different
instantiations, although they are qualitatively similar.
For $N$$=$$24$ this variation disappears;
this is a result of the well-studied self-averaging property of the SYK model \cite{garcia2016spectral,Cotler2017}. We 
take advantage of this property in our analysis to construct and study 
examples using a single instantiation of the SYK couplings .

\section{Warm up exercise: the Gao Jafferis protocol with 
{\boldmath $t_0$$=$$t_1$$=$$\beta$$=$$0$}}\label{sec:warmup}

For $t_0$$=$$t_1$$=$$\beta$$=$$0$ we compute the reduced density
matrix for the reference and readout qubits $RT$
explicitly as a function of the interaction strength $\mu$. Starting from
eqn. \ref{eq:rhoRTdef}, and using the identities given in Appendix A,
the result is:
\begin{align}
\label{eq:rhoRTfinal}
\rho_{TR}(t=0, \beta = 0) = \frac{1}{2}
\begin{pmatrix}
\frac{1}{2}(1+{\rm sin}^2\mu )& 0 & 0 & {\rm sin}\mu \\
0 & \frac{1}{2}(1-{\rm sin}^2\mu )& 0 & 0 \\
0 & 0 & \frac{1}{2}(1-{\rm sin}^2\mu )& 0 \\
{\rm sin}\mu & 0 & 0 & \frac{1}{2}(1+{\rm sin}^2\mu )
\end{pmatrix}
\end{align}
from which the further reduced density matrices of $R$ and $T$ follow:
\begin{align}
\label{eq:rhoRandT}
\rho_R(t=0, \beta = 0) = \rho_T(t=0, \beta = 0) = \frac{1}{2}
\begin{pmatrix}
1 & 0 \\
0 & 1
\end{pmatrix}
\end{align}

We compute the corresponding mutual information $I(R$$:$$T)$ using the
second Renyi entropies:
\begin{align}
\label{eq:IRT}
I(R:T) &= S(R) + S(T) - S(RT) \nonumber\\
&= {\rm log}_2\left( \frac{{\rm Tr}(\rho_{RT}^2)}
{{\rm Tr}(\rho_R^2){\rm Tr}(\rho_T^2)}\right) \nonumber\\
&= 2\,{\rm log}_2 \left( 1 + {\rm sin}^2\mu \right)
\end{align}
We find that $I(R$$:$$T)$ vanishes for $\mu$$=$$0$ (i.e. no Left-Right interaction), and
is maximal, $I(R$$:$$T)$$=$$2$ in units of log(2), for $|\mu|$$=$$\pi/2$.
This phenomenon has nothing to do with wormhole teleportation or any
other kind of teleportation, as we demonstrate below.

The Left-Right interaction is written explicitly as a unitary gate
operation on qubits. We represent the Left Majoranas
$\psi_0^\ell$, and $\psi_1^\ell$ as acting on the same left qubit $\ell$,
and the Right Majoranas $\psi_0^r$, and $\psi_1^r$ as acting on the 
same Right qubit $r$, as the following Pauli matrix operators:
\begin{align}
\label{eq:qubitops}
\psi_0^\ell &= \frac{1}{\sqrt{2}}X^\ell \; ;\quad\quad\quad
\psi_1^\ell = \frac{1}{\sqrt{2}}Y^\ell \nonumber\\
\psi_0^r &= \frac{1}{\sqrt{2}}Z^\ell X^r \; ;\quad\quad
\psi_1^r = \frac{1}{\sqrt{2}}Z^\ell Y^r 
\end{align}
where the Pauli $Z$ matrices are needed, as per Jordan-Wigner, to enforce anti-commutation.

It follows that 
\begin{align}
\label{eq:V01}
V_{01} = \frac{1}{2}\left( Y^\ell X^r - X^\ell Y^r \right)
\end{align}
so
\begin{align}
\label{eq:V01c}
{\rm e}^{i\mu V_{01}} = \left( {\rm cos}\frac{\mu}{2}\,{\cal I}_4 
+ i\,{\rm sin}\frac{\mu}{2}\,Y^\ell X^r \right)
\left( {\rm cos}\frac{\mu}{2}\,{\cal I}_4 
- i\,{\rm sin}\frac{\mu}{2}\,X^\ell Y^r \right)
\end{align}
where ${\cal I}_4$ is the 4d identity matrix. We can write this more explicitly 
as the following 2-qubit unitary gate operation:
\begin{align}
\label{eq:V01matrix}
{\rm e}^{i\mu V_{01}} =
\begin{pmatrix}
1 & 0 & 0 & 0 \\
0 & {\rm cos}\mu & {\rm sin}\mu & 0 \\
0 & -{\rm sin}\mu & {\rm cos}\mu & 0 \\
0 & 0 & 0 & 1
\end{pmatrix}
\end{align}
It is instructive to rewrite this as a product of two unitary operators:
\begin{align}
\label{eq:V01matrixprod}
{\rm e}^{i\mu V_{01}} =
\begin{pmatrix}
1 & 0 & 0 & 0 \\
0 & 1 & 0 & 0 \\
0 & 0 & -1 & 0 \\
0 & 0 & 0 & 1
\end{pmatrix}
\cdot
\begin{pmatrix}
1 & 0 & 0 & 0 \\
0 & {\rm cos}\mu & {\rm sin}\mu & 0 \\
0 & {\rm sin}\mu & -{\rm cos}\mu & 0 \\
0 & 0 & 0 & 1
\end{pmatrix}
\end{align}
The reader can easily check that the first matrix is the
2-qubit entangling operation CNOT$\cdot$CZ$\cdot$CNOT,
with the Left qubit $\ell$ as the control.
The second matrix, for $\mu$$=$$\pi/2$, is the canonical 2-qubit SWAP gate:
\begin{align}
\label{eq:V01matrixc}
{\rm SWAP} =
\begin{pmatrix}
1 & 0 & 0 & 0 \\
0 & 0 & 1 & 0 \\
0 & 1 & 0 & 0 \\
0 & 0 & 0 & 1
\end{pmatrix}
\end{align}
The presence of swapping implies that this transfer of quantum
information cannot be interpreted as teleportation: the 
deferred measurement principle does not apply to swapping; this is easily seen 
explicitly by writing the 
SWAP operator as three successive CNOTs with control and target qubits alternating.
Thus we cannot use a classical
channel to reproduce the same effect. 

\begin{figure}[htb]
\begin{center}
\includegraphics[width=0.45\textwidth]{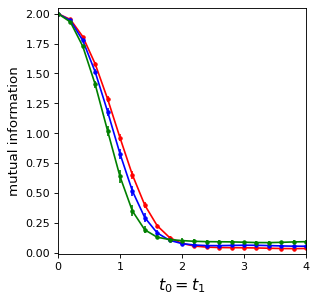}
\protect\caption{Time dependence of the direct
transmission mechanism. Mutual information $I(R$$:$$T)$ after
running the wormhole teleportation protocol with injection at time
$-t_0$ and extraction at time $t_1$$=$$t_0$. The red/blue/green points are results averaging 10 
instantiations of the $N$$=$$10$ SYK model with Left-Right interaction $V$,
using $\cal J$$=$$1$,
and $\beta$ = 0,4, and 16 respectively.
We have chosen $|\mu |$$=$$\pi/2$ so that the direct transmission
has perfect fidelity for $t_0$$=$$t_1$$=$$0$.
\label{fig:comparetzeroplotsb}}
\end{center}
\end{figure}

Even in this simple example with $\mu$$=$$\pi/2$
we also see a more general puzzle related to
implementing a classical channel, arising from the Pauli $Z$ operations
needed to enforce anti-commutation of the Right Majoranas with the Left Majoranas.
These imply that the operator $S_{Tr}$ is not just the canonical 2-qubit SWAP gate
applied to qubits $T$ and $r$, but also acts on the Left qubit $\ell$.
\begin{align}
\label{eq:actual_swap_a}
{\rm S_{Tr}} = {\rm CZ}(T\to\ell )\cdot {\rm CZ}(r\to\ell ) \cdot {\rm SWAP}(Tr)
\end{align}
where the direction of the arrows in the CZ gate go from the control qubit
to the target qubit. This feature, which will appear in any implementation
of wormhole teleportation using fermions, appears at first glance to be an obstacle
to implementing a classical channel. Fortunately we can use the fact
that for any pair of qubits $a$,$b$, CZ($a$$\to$$b$) = CZ($b$$\to$$a$), thus
\begin{align}
\label{eq:actual_swap_b}
{\rm S_{Tr}} = {\rm CZ}(\ell\to T )\cdot {\rm CZ}(\ell\to r ) \cdot {\rm SWAP}(Tr)
\end{align}
This means, again appealing to the deferred measurement principle, that the
CZ gates can be replaced by a classical channel between Alice and Bob. In fact,
only the first CZ gate has an effect on $I(R$$:$$T)$; the second one can be dropped.

As a final insight from this example, we study the case of
$\mu$$=$$\pi/2$ when we go to finite temperature and 
some range of nonzero time evolution. 
Figure \ref{fig:comparetzeroplotsb} shows results using the Left-Right interaction $V$ and three different
values of $\beta$: 0, 4, and 16. 
We use $|\mu |$$=$$\pi/2$ so that the direct transmission
has perfect fidelity for $t_0$$=$$t_1$$=$$0$ and thus dominates over
any teleportation effect at early times.
We see that the direct transmission effect 
drops off rapidly with time, becoming negligible for $t \gtrsim 2$, although
a small residual remains. The rapid falloff is due to the efficient scrambling from the SYK Hamiltonian. As expected, the characteristic time scale for the
swapping effect to become negligible is roughly the same as the timescale for
the wormhole teleportation to turn on.

\section{Long-range protocol}\label{sec:lrprotocol}

Consider replacing the Left-Right interaction 
exp$(i\mu V)$ with exp$(i\mu V^b)$, where $V^b$ is defined by
\begin{align}
\label{eq:Vbdef}
V^b  = 
\sum_{j=0}^{N/2-1} \, 
\Gamma^{(2)\ell}_j \Gamma^{(2)r}_j 
\end{align}
where
\begin{align}
\label{eq:Gammadef}
\Gamma^{(2)\ell}_j  = 2i\psi^\ell_{2j}\psi^\ell_{2j+1}\; ; \quad
\Gamma^{(2)r}_j  = 2i\psi^r_{2j}\psi^r_{2j+1}
\end{align}

We can give an alternate expression for $V^b$ by substituting
a Jordan-Wigner representation on $N$ qubits for the $2N$ Majoranas:
\begin{align}
\label{eq:JW}
\psi_{2j}^\ell &= \frac{1}{\sqrt{2}}Z_0\ldots Z_{j-1}
X_j \; ;\quad
\psi_{2j+1}^\ell = \frac{1}{\sqrt{2}}Z_0\ldots Z_{j-1}
Y_j \; ;\quad j = 0,1,\ldots N/2-1\nonumber\\
\psi_{2j}^r &= \frac{1}{\sqrt{2}}Z_0\ldots Z_{j-1+N/2}
X_{j+N/2} \; ;\quad
\psi_{2j+1}^r = \frac{1}{\sqrt{2}}Z_0\ldots Z_{j-1+N/2}
Y_{j+N/2} 
\end{align}
From which we see that:
\begin{align}
\label{eq:ZZdef}
V^b  = 
\sum_{j=0}^{N/2-1} \, 
Z_{j}Z_{j+N/2}
\end{align}
Eqn. \ref{eq:ZZdef} is the expression proposed originally by
\citet{nezami2021quantum}.

\subsection{Absence of direct transmission in the long-range
protocol}

If we replace $V$ by $V^b$,
the computation of the matrix elements in equations \ref{eq:rhodefsa} becomes trivial. This is because
of the identities:
\begin{align}
\label{eq:V01b}
{\rm e}^{-i\mu V_{01}^b} \,\Gamma_0^{(2)r} \,{\rm e}^{i\mu V_{01}^b} = \Gamma_0^{(2)r} \; ;\quad\quad
{\rm e}^{-i\mu V_{01}^b} \,\chi_r^\dagger \,{\rm e}^{i\mu V_{01}^b} = -\chi_r^\dagger
\end{align}
from which we derive:
\begin{align}
\label{eq:rhoRTbosonic}
\rho_{RT}(t=0, \beta = 0) = \frac{1}{2}
\begin{pmatrix}
\frac{1}{2} & 0 & 0 & 0 \\
0 & \frac{1}{2}& 0 & 0 \\
0 & 0 & \frac{1}{2}& 0 \\
0 & 0 & 0 & \frac{1}{2} 
\end{pmatrix}
\end{align}
which implies $I_{RT}(t$$=$$0, \beta$$=$$0) = 0$.

Another way to see this is to observe that:
\begin{align}
\label{eq:V01bexpand}
{\rm e}^{i\mu V_{01}^b} &= {\rm cos}\mu\,{\cal I}_4 + i\,{\rm sin}\mu\,\Gamma_0^{(2)\ell}\Gamma_0^{(2)r} 
\nonumber\\
&= {\rm cos}\mu\,{\cal I}_4 + i\,{\rm sin}\mu\,Z^\ell Z^r \\
&= 
\begin{pmatrix}
{\rm e}^{i\mu} & 0 & 0 & 0 \\
0 & {\rm e}^{-i\mu}& 0 & 0 \\
0 & 0 & {\rm e}^{-i\mu} & 0 \\
0 & 0 & 0 & {\rm e}^{i\mu} 
\end{pmatrix} \nonumber
\end{align}
which, unlike the expression in eqn \ref{eq:V01matrix}, is obviously not a Left-Right swap.

\begin{figure}[htb]
\begin{center}
\includegraphics[width=0.43\textwidth]{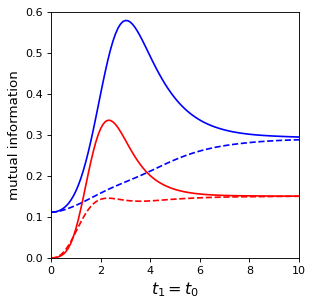}
\includegraphics[width=0.43\textwidth]{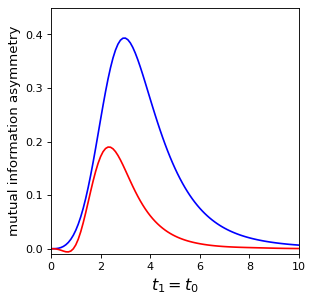}
\protect\caption{{\bf Left:} mutual information (computed using the second Renyi entropies) 
between the qubits $R$ and $T$ after
running the wormhole teleportation protocol with injection at time
$-t_0$ and extraction at time $t_1$$=$$t_0$. Results from a single
instantiation of the $N$$=$$24$, $q$$=$$4$, SYK model with $\beta$$=$$4$,
$\cal J$$=$$1$, using either the Left-Right interaction
$V$ (blue lines) with $\mu $$=$$-0.2$ (solid)/$\mu $$=$$+0.2$ (dashed), or $V^b$ (red lines) with the same values of $\mu$.
{\bf Right:} the asymmetry in the $RT$ mutual information between using negative/positive values of the coupling $\mu$ in the protocol.
\label{fig:compareVvsVb}}
\end{center}
\vspace{-0.4cm}
\end{figure}

\subsection{Quantum information transfer with the long-range
protocol}

Figure \ref{fig:compareVvsVb} compares the results of the 
original wormhole teleportation protocol using $V$ and the
modified protocol using $V^b$. In both cases we take
$N$$=$$24$, $q$$=$$4$, $\beta$$=$$4$, ${\cal J}$$=$$1$,
and $\mu = \pm 0.2$. As previously remarked, using $V^b$ there is
no direct transmission, so the mutual information vanishes
at $t_0$$=$$0$. Instead, as $t_0$ is increased, we observe
a smooth ramp up of the holographic contribution,
leading to a mutual information peak only for negative $\mu$, and the non-holographic contribution from  the ``peaked-size teleporation'',
that approaches the same late time asymptotic value for
either sign of $\mu$. The fact that the mutual information
between qubits $R$ and $T$ develops a substantial time-dependent
asymmetry, depending on whether the sign of $\mu$ corresponds to
a negative versus positive energy pulse in the holographic picture,
is a signature feature of wormhole teleportation, and as we
demonstrate in section \ref{sec:sizewinding} agrees 
with the time evolution of the size winding.
The overall transfer of quantum information
is less efficient using $V^b$ rather than $V$; this is consistent with
the differences in the efficiency of the size winding
shown in the next section.

\section{Size winding}\label{sec:sizewinding}
Size winding \cite{brown2021quantum,nezami2021quantum}
is a key feature of wormhole teleportation. It
directly relates a coherent property of the operator spreading induced by SYK scrambling
dynamics to motion along the emergent light cone through the traversable wormhole 
\cite{maldacena2017diving},\cite{Maldacena:2018lmt},\cite{Lin:2019qwu}. The holographic duality
of wormhole teleportation is precisely the
statement that the complete size winding description and the
through-the-wormhole description have a 
mapping into each other. In this sense it is reasonable to
use the size winding description {\it as our definition}
of what we mean by traversable wormhole behavior
away from semi-classical regime. Note, however,
that the presence of individual components of the size winding description
does not necessarily imply a holographic dual.

\subsection{Review of size winding with the Gao-Jafferis protocol}\label{sec:revew-sw}
We briefly review the size winding 
mechanism and demonstrate how it works  in simulations using the original Gao-Jafferis protocol with Left-Right interaction $V$.

A single Majorana fermion $\psi_i^\ell$,
$\psi_i^r$ in the Left or Right SYK systems will
time evolve as
\begin{align}
\label{eq:timeevolve}
\psi_i^\ell (t) = {\rm e}^{iH_Lt} \psi_i^\ell {\rm e}^{-iH_Lt}\; ;
\quad
\psi_i^r (t) = {\rm e}^{iH_Rt} \psi_i^r {\rm e}^{-iH_Rt}\
\end{align}
We can expand either of these in the terms of the
the complete set of left/right basis operators $\Gamma_J^{(s)\ell}$,$\Gamma_J^{(s)r}$
as defined, e.g. in
\cite{Qi2019,Gao:2023gta}:
\begin{align}
\label{eq:basisdef}
\Gamma^{(s)}_J  = 2^{s/2}i^{s(s-1)/2}\psi_{j_1}\ldots\psi_{j_s}
\end{align}
where $J$ represents the string of ordered indices
$j_1$,$j_2$,$\dots j_s$ labeling some combination of $s$ Majoranas. 
These basis operators are Hermitian,
square to the identity,
and create mutually orthogonal states acting on $|I\rangle$.
Note also the relation:
\begin{align}
\label{eq:basisrelation}
\Gamma^{(s)\ell}_J \ket{I} = (-i)^s \Gamma^{(s)r}_J \ket{I}
\end{align}

The scrambling dynamics of the SYK model imply that
over time $\psi_i^\ell (t)$ and $\psi_i^r (t)$ will
have support on a larger and larger number of basis
operators. Since each basis operator is a string of $s$
Majoranas, the effective size of $\psi_i^\ell (t)$ and $\psi_i^r (t)$
will grow over time. To study this in more detail,
it is useful to follow \citet{Qi2019} and do the same expansion for the thermal fermion operators:
\begin{align}
\label{eq:opexpand}
O_L^\beta(t) &\equiv \psi_i^\ell (t)\rho^{1/2}_\beta = \frac{1}{\sqrt{2}}
\sum_I c_I^\ell (t) \Gamma_J^{(s)\ell} \nonumber\\
O_R^\beta(t) &\equiv \psi_i^r (t)\rho^{1/2}_\beta = \frac{1}{\sqrt{2}}
\sum_I c_I^r (t) \Gamma_J^{(s)r}
\end{align}
where:
\begin{align}
\label{eq:opexpandcoeff}
c_J^\ell (t) &= \bra{I} \Gamma_J^{(s)\ell}\, \psi_i^\ell (t) \ket{\rm tfd}
\nonumber\\
c_J^r (t) &= \bra{I} \Gamma_J^{(s)r}\, \psi_i^r(t) \ket{\rm tfd}
\end{align}
The size distribution of the operators as a function of time
can then be computed from:
\begin{align}
\label{eq:Pdef}
P^\ell(s) &= \sum_{|J| = s} | c^\ell_J(t) |^2 \nonumber\\
P^r(s) &= \sum_{|J| = s} | c^r_J(t) |^2
\end{align}

\begin{figure}[htb]
\begin{center}
\includegraphics[width=0.45\textwidth]{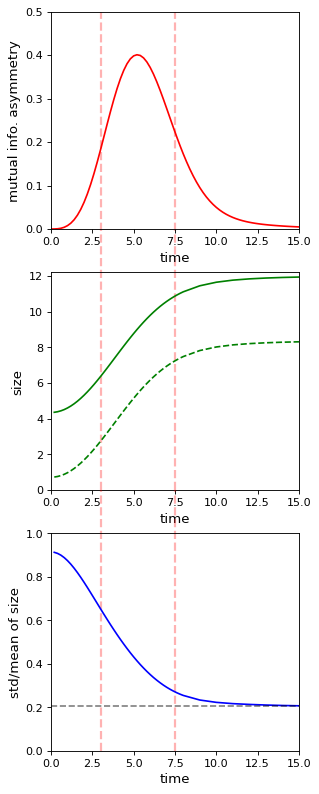}
\protect\caption{Operator size growth for the wormhole
teleportation protocol with $N$$=$$24$, $q$$=$$8$, ${\cal J}$$=$$1$,
$\beta$$=$$8$, and $\mu$$=\pm 0.2$. {\bf Top:} The mutual information asymmetry as a function of time. {\bf Middle:} The mean size of the
thermal fermion operator $\psi_i^\ell (t)\rho^{1/2}$ (solid line)
and the same quantity but subtracting off the size of $\rho^{1/2}$
(dashed line). {\bf Bottom:} The standard deviation of the thermal fermion
size distribution divided by the mean size (solid line); for
comparison the dashed line shows $1/\sqrt{N}$.
\label{fig:sizegrowth}}
\end{center}
\end{figure}

Figure \ref{fig:sizegrowth} shows the operator size growth over
time of the thermal fermion $O_L^\beta(t)$, for the case
$N$$=$$24$, $q$$=$$8$, ${\cal J}$$=$$1$, $\beta$$=$$8$.
For comparison, the top plot shows the mutual information asymmetry,
i.e. the difference between $I(R$$:$$T)$ resulting from performing the
teleportation protocol using a negative
coupling $\mu$$=- 0.2$ and $I(R$$:$$T)$ resulting from using a positive
coupling $\mu$$=+ 0.2$. The vertical dashed red line lines
define the approximate range of injection times (enforcing, for simplicity, $t_0$$=$$t_1$) over which the holographic dynamics
dominates the transfer of quantum information. 

The solid line in the middle plot shows the mean size of a
left thermal fermion, which we denote as 
$n[\psi_i^\ell (t)\rho^{1/2}]$; the dashed line shows the
difference between this quantity and the time-independent
size of the thermal factor $\rho^{1/2}_\beta$:
\begin{align}
\label{eq:rhosizedef}
n[\rho^{1/2}_\beta] = \frac{N}{2}\left( 1 - G(\frac{\beta}{2})\right)
\end{align}
where $G(\frac{\beta}{2})$ is the Euclidean two-point function of
that thermal fermion \cite{Qi2019}:
\begin{align}
\label{eq:Gdef}
G(\frac{\beta}{2}) = 2Z^{-1}_\beta {\rm Tr}\left(
{\rm e}^{-\beta H_L}\psi^\ell_i(\frac{\beta}{2})\psi^\ell_i(0) \right)
= 2\bra{I} \psi_i^\ell \rho^{1/2}_\beta
\psi_i^\ell \rho^{1/2}_\beta \ket{I}
\end{align}
Note $G(\frac{\beta}{2})$ varies from near one at small $\beta$ to
near zero at large $\beta$.

The middle plot shows that the operator size
is growing rapidly over the range of times where the 
holographic teleportation is effective.
In the large $N$, large $q$ limit this
growth is exponential, and the exponent is called the Lyapunov
exponent. In this limit one has \cite{Qi2019}:
\begin{align}
\label{eq:growthinlimit}
n[\psi_i^\ell (t)\rho^{1/2}_\beta] - n[\rho^{1/2}_\beta]
 = G(\frac{\beta}{2})\left( 1 + \frac{8{\cal J}^2}{\lambda^2}
{\rm sinh}^2\frac{\lambda t}{2} \right)
\end{align}
where $\lambda$ denotes the Lyapunov exponent. 
This exponent has an upper bound, $\lambda \leq 2\pi/\beta$;
black holes saturate this upper bound, as does the SYK model in
the limit $N$$\to$$\infty$,  $\beta{\cal J}$$\to$$\infty$, 
$N/\beta{\cal J}$$\to$$\infty$.
While the growth described by
eqn. \ref{eq:growthinlimit} is exponential, for 
$\lambda t/2 \lessapprox 1$ it reduces to a quadratic in time with
the $\lambda$ dependence dropping out. This is the regime that
we are in for the example of Figure \ref{fig:sizegrowth}, so it is
not useful to try to extract the Lyapunov exponent directly from
the operator size growth in this regime of parameters.

To study size winding, we begin by observing that,
using eqns. \ref{eq:Istate}, \ref{eq:basisrelation} and the 
hermiticity of the basis operators, it is straightforward
to show:
\begin{align}
\label{eq:coeffrelation}
\left( c^{r*}_J(t) \right)^2 = \left( c^\ell_J(-t) \right)^2
\end{align}
Size winding is observed using the distributions:
\begin{align}
\label{eq:Qdef}
Q^\ell(s) = \sum_{|J| = s} \left( c^\ell_J(t) \right)^2 \; ;\quad
Q^r(s) = \sum_{|J| = s} \left( c^r_J(t) \right)^2
\end{align}
The size winding ansatz of \citet{brown2021quantum,nezami2021quantum} is:
\begin{align}
\label{eq:swansatz}
Q^\ell(t) = {\rm e}^{i\alpha s/N}r^2, 
\; ;\quad
Q^r(t) = {\rm e}^{-i\alpha s/N}r^2 \quad \alpha,r \in \mathbb{R}
\end{align}
This is a strong ansatz since it assumes both a linear
dependence of the phase on size, and an overall coherence involving
the large number of distinct basis operators
$\Gamma^{(s)\ell}_J$, $\Gamma^{(s)r}_J$ of a given size.

Notice that the phases of the individual $c_J$ depend on the
choice of basis operators $\Gamma^{(s)}_J$; if we rotate the basis
operators of a given size $s$ by a real orthogonal matrix $R^{(s)}_{JK}$,
the $c_J$ change but the corresponding $Q(s)$ are basis independent:
\begin{align}
\label{eq:basis_change}
\tilde{\Gamma}^{(s)}_J \equiv \sum_K \Gamma^{(s)}_K R_{KJ} \; ;\quad
\tilde{c}^{(s)}_J \equiv \sum_K c^{(s)}_K R_{KJ} \nonumber\\
\sum_J c^{(s)}_J \Gamma^{(s)}_J 
= \sum_J \tilde{c}^{(s)}_J \tilde{\Gamma}^{(s)}_J \; ;\quad
\sum_J \left( c^{(s)}_J \right)^2 
= \sum_J \left( \tilde{c}^{(s)}_J \right)^2 
\end{align}

As in \cite{brown2021quantum,nezami2021quantum}, we define ``perfect size winding'' by the requirement $|Q(s)|/P(s) = 1$
for all sizes $s$. This would require that all the individual $c_J$
for a given size have the same phase, which in this limit is basis-independent
since an overall phase factors out in the expressions above. Since
the size winding in the systems we analyze  is not perfect, we 
confine our attention to the basis independent quantities $Q(s)$. However we use the
ratio $|Q(s)|/P(s)$ as a diagnostic of the overall
coherence of the size winding phenomenon.

To understand the effect of the Left-Right interaction operator
exp$i\mu V$ on size winding, let's rewrite the expression
eqn. \ref{eq:Vdef} for $V$ as:
\begin{align}
\label{eq:Vdef2}
V  = \frac{i}{2}
\sum_{j=0}^{N-1} \, 
\Gamma_j^{(1)\ell}\Gamma_j^{(1)r}
\end{align}
where we have used the size 1 basis operators
$\Gamma_j^{(1)\ell} = \sqrt{2}\psi_j^\ell$ and
$\Gamma_j^{(1)r} = \sqrt{2}\psi_j^r$, which obey the
following commutation relations with other left basis operators:
\begin{align}
\label{eq:Gamma1identities}
\Gamma_j^{(1)\ell}\Gamma_j^{(1)r}\Gamma_J^{(s)\ell}
=(-1)^{j\cap J} \Gamma_J^{(s)\ell} \Gamma_j^{(1)\ell}\Gamma_j^{(1)r}
\end{align}
Given also that $\Gamma_j^{(1)\ell}\Gamma_j^{(1)r}\ket{I} = i\ket{I}$,
this implies
\begin{align}
\label{eq:sizeop}
\bra{I}\Gamma_J^{(s)\ell}\,V\,\Gamma_J^{(s)\ell}\ket{I}
&= -\frac{1}{2}\sum_{|j\cap J|=0} + \frac{1}{2}\sum_{|j\cap J|=1}
\nonumber\\
&= s - \frac{N}{2}
\end{align}
Thus $V$, up to an overall additive constant, measures the
size $s$ of any basis operator, and the Left-Right interaction
exp$i\mu V$ applies a size-dependent phase that corresponds to
shifting $Q^\ell (s)$ by $2\mu s$. 

What does this imply for the wormhole teleportation protocol?
If at injection time $c^\ell_J(-t_0)$ already possesses
the linear size dependence of eqn. \ref{eq:swansatz} with some
positive slope $\alpha/N$, then applying the Left-Right interaction
with $\mu \simeq -\alpha/N$ at $t=0$ can have the effect of
reversing the sign of the size winding, thus
coherently converting Left side size winding to Right
side size winding.

The holographic description of teleportation through the wormhole
with two entangled copies of the SYK model is explained in detail in
\cite{maldacena2017diving,Maldacena:2018lmt,Lin:2019qwu}, and the mapping between this
description and size winding is discussed in
\cite{brown2021quantum,nezami2021quantum}.
In the holographic picture an emergent spatial degree of freedom combines
with time in a nearly-AdS$_2$ geometry; the geometry is nearly-AdS in that it includes leading effects that break conformal symmetry,
consistent with the fact that the low energy correlators of the SYK model are ``nearly'' conformal. 
Gravity in such a system was first described by Jackiw and
Teitelboim \cite{Jackiw:1984je,Teitelboim:1983ux}; it has no propagating
degrees of freedom, but there is a gravitational mode that maps the proper time
on the nearly-AdS boundary to AdS$_2$ time as expressed in
Poincare or Rindler coordinates.
The holographic dual system of JT gravity coupled to bulk matter
has an approximate $SL(2,R)$ symmetry generated by $B$, $E$, and
$-i[B,E]$, where
\begin{align}
\label{eq:SL2R_gen}
B = H_R - H_L\; ;\quad
E = H_L + H_R - \mu V - E_0
\end{align}
and $E_0$ is a constant.
The boost operator $B$ annihilates our $t$$=$$0$ state, i.e., the thermofield double state $\ket{\rm tfd}$. The operator $E$, which can be considered as
generating global time translations, also approximately annihilates
$\ket{\rm tfd}$ for a suitable choice of $E_0$; we  quantify this
statement with examples in subsection \ref{sec:eternal}.
Translations along the two bulk null directions are generated by
\begin{align}
\label{eq:Pnull}
P_\pm = -\frac{1}{2}\left( E \pm B \right)
\end{align}
From this correspondence it was shown in \cite{brown2021quantum,nezami2021quantum}
that size winding describes the momentum wavefunction of 
a particle traversing the emergent wormhole, where the conjugate position
describes the particle location relative to the horizon.

\vspace{-0.3cm}
\begin{figure}[htb]
\begin{center}
\includegraphics[width=0.4\textwidth]{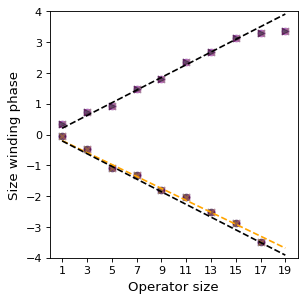}
\includegraphics[width=0.4\textwidth]{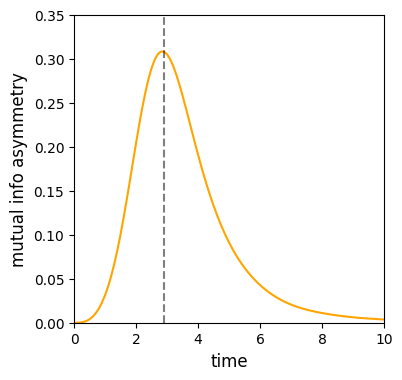}
\protect\caption{Example of size winding. {\bf Left:} Phase of $Q^\ell(s)$ as a function of the operator size $s$ for the first 10 fermions of a single
instantiation of the wormhole protocol with $N$$=$$20$, $q$$=$$4$, $\beta$$=$$4$, ${\cal J}$$=$$1$, $t_0$$=$$2.9$,
$\mu$$=$$-0.2$, using $V$ to generate the
Left-Right interaction; the upper dashed black line is a linear fit to the Left size winding, weighted by the probability $P^\ell(s)$;
the lower dashed black line is the corresponding Right size winding, and
the orange dashed line is a linear fit to the actual
points after applying the Left-Right interaction. {\bf Right:} The mutual information asymmetry for the same
parameters, fixing also $t_1$$=$$t_0$; the dashed line shows
$t_0$$=$$2.9$.
\label{fig:q4t29comp}}
\end{center}
\vspace{-0.3cm}
\end{figure}

Figure \ref{fig:q4t29comp} (Left) shows size winding for the case
$N$$=$$20$, $q$$=$$4$, $\beta$$=$$4$, ${\cal J}$$=$$1$, $t_0$$=$2.9, $\mu$$=$$-0.2$, using $V$ for the Left-Right
interaction.
The time chosen
corresponds approximately to the peak in the mutual information
asymmetry. We see excellent agreement with the expected linear
dependence of the phase on size, and that the Left-Right interaction flips Left size winding to Right size winding, in agreement
with the holographic interpretation with a traversable wormhole.
The figure superimposes the phases for 10 of the 20 Majorana fermions
in the model; the fact that every fermion has identical
behavior is a consequence of the self-averaging of the SYK model
for this value of $N$.

\begin{figure}[htb]
\vspace{-0.5 cm}
\begin{center}
\includegraphics[width=0.41\textwidth]{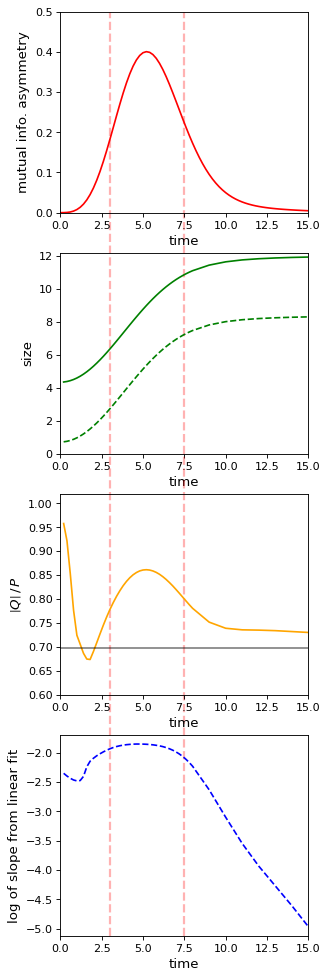}
\protect\caption{Size winding of the Gao-Jafferis wormhole
teleportation protocol with $N$$=$$24$, $q$$=$$8$, ${\cal J}$$=$$1$,
$\beta$$=$$8$, and $\mu$$=\pm 0.2$. {\bf Top:} The mutual information asymmetry as a function of time. {\bf Second:} The mean size of the
thermal fermion operator. {\bf Third:} $|Q|/P$ (orange line) averaged over size;
the solid black line denotes $G(\beta )$.
{\bf Bottom:} The log of the fitted linear size winding slope.
\label{fig:all}}
\end{center}
\end{figure}

\begin{figure}[htb]
\begin{center}
\includegraphics[width=0.4\textwidth]{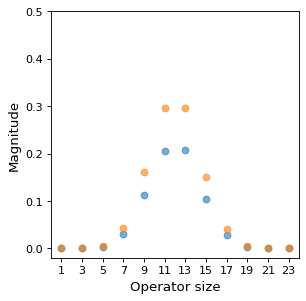}
\includegraphics[width=0.4\textwidth]{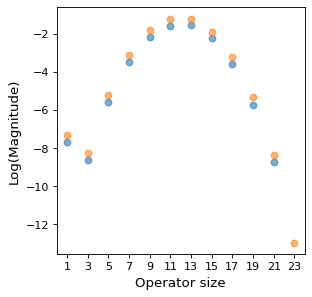}
\protect\caption{Distribution of the thermal fermion operator
size at the scrambling time $t$$=$$15$, for the
$N$$=$$24$ SYK model with $q$$=$$8$, $\cal J$$=$$1$, $\beta$$=$$8$,
shown on a linear scale (left) and a log scale (right).
The orange points are $P(s)$ as defined in eqn. \ref{eq:Pdef};
also shown (blue points) are $|Q(s)|$ as defined in
eqn. \ref{eq:Qdef}.
\label{fig:latetime}}
\end{center}
\vspace{-0.4cm}
\end{figure}

\begin{figure}[htb]
\begin{center}
\includegraphics[width=0.4\textwidth]{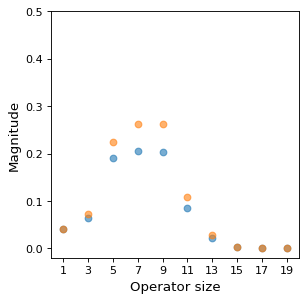}
\includegraphics[width=0.4\textwidth]{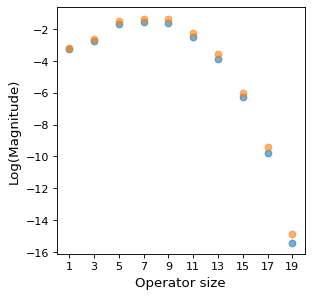}
\protect\caption{Distribution of the thermal fermion operator
size at time $t$$=$$2.9$. Shown are 
$|Q^\ell(s)|$ (blue) and
$P^\ell(s)$ (orange) as a function of the operator size $s$, averaged over the first 10 fermions of a single
instantiation of the wormhole teleportation protocol with $N$$=$$20$, $q$$=$$4$, $\beta$$=$$4$, ${\cal J}$$=$$1$,
on a linear scale (left) and a log scale (right).
\label{fig:q4t29sizes}}
\end{center}
\vspace{-0.4cm}
\end{figure}

\subsection{Peaked-size teleportation}\label{sec:peaksize}

A universal feature of SYK scrambling
dynamics is that, as the teleportation time interval is increased,
the mean operator size eventually asymptotes to $N/2$. This scrambling time is roughly 15 for the example shown 
in Figure \ref{fig:sizegrowth}. As seen in Figure \ref{fig:latetime},
at the scrambling time the operator size distribution is strongly and symmetrically peaked around $N/2$. The lower plot of
Figure \ref{fig:sizegrowth} shows that the standard deviation of
the operator size distribution, in units of the mean size,
starts out close to one at early times, but by the scrambling time
has asymptoted to $1/\sqrt{N}$. 

As observed in \cite{schuster2021manybody}, if we choose the injection
$t_0$ in the teleportation protocol to be greater than 
or equal to the scrambling time,
then at $t$$=$$0$ the size of a thermal fermion for large $N$ will
be strongly peaked at $N/2$. This allows the Left-Right interaction using the size operator to preserve quantum
information transfer without invoking size winding. 

This 
non-holographic many-body quantum effect, dubbed peaked-size teleportation  in \cite{schuster2021manybody},
will also be present at some level for $t_0$ values smaller
then the scrambling time, including values where our through-the-wormhole quantum teleportation dominates. This is quantified in the long-range wormhole teleportation protocol, where the
direct transmission mechanism is absent. Thus for example the
red dashed line in Figure \ref{fig:compareVvsVb} represents
quantum information transfer coming entirely from the ``peaked-size teleportation''.  In the infinite temperature limit (i.e. $\beta \to 0$)
the through-the-wormhole teleportation disappears while the
peaked-size teleporation remains; the size winding phases also
disappear in this limit, since for $\beta$$=$$0$ the operators
in Eq. \ref{eq:opexpand} are Hermitian.

Figure \ref{fig:q4t29sizes} shows the thermal fermion size distribution
in an example of the Gao-Jafferis wormhole teleportation with $t_0$$=$$t_1$$=$$2.9$,
approximately the choice that optimizes the holographic
quantum information transfer. Note that the size distribution is not
centered around $N/2$ and is more spread out than the late time example of Figure \ref{fig:latetime}.

\subsection{Size winding with the long-range protocol}\label{sec:Vb-sw}

The alert reader may at this point despair for the
long-range protocol, in which $V$ has been replaced by $V^b$.
How can we preserve the holographic size winding feature 
without using the size operator $V$? The answer is that $V^b$ is also
a size operator equivalent to $V$.
To see this, start with the definition of $V^b$ in
eqn. \ref{eq:Vbdef} and write the commutation relations
analogous to eqn. \ref{eq:Gamma1identities}:
\begin{align}
\label{eq:Gamma2identities}
\Gamma_j^{(2)\ell}\Gamma_j^{(2)r}\Gamma_J^{(s)\ell}
=(-1)^{j\cap J} \Gamma_J^{(s)\ell} \Gamma_j^{(2)\ell}\Gamma_j^{(2)r}
\end{align}
Then we use $\Gamma_j^{(2)\ell}\Gamma_j^{(2)r}\ket{I} = -\ket{I}$
to find:
\begin{align}
\label{eq:size2op}
\bra{I}\Gamma_J^{(s)\ell}\,V^b\,\Gamma_J^{(s)\ell}\ket{I}
&= -\sum_{|j\cap J|=0} + \sum_{|j\cap J|=1}
- \sum_{|j\cap J|=2}
\nonumber\\
&= s_2 - \frac{N}{2}
\end{align}
where $s_2$ is the size measured by $V^b$.
In terms of Majorana operator strings realized on qubits, it is easy to see the relation between the size
$s$ measured by $V$ and the size $s_2$ measured by $V^b$:
$s$ counts the number of Majoranas in the operator string,
while $s_2$ counts the number of Majoranas in the operator string
not counting pairs of Majoranas residing on the
same qubit. Note that the maximum value of $s_2$ is $N/2$.

\begin{figure}[htb]
\begin{center}
\includegraphics[width=0.41\textwidth]{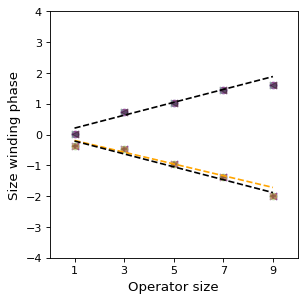}
\includegraphics[width=0.41\textwidth]{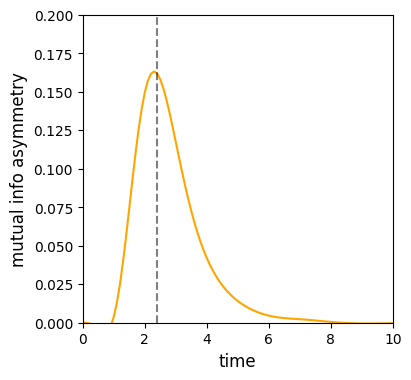}
\protect\caption{Example of size winding in the long-range wormhole
teleportation protocol.
{\bf Left:} Phase of $Q^\ell(s)$ as a function of the operator size $s$ for the first 10 fermions of a single
instantiation of the long-range wormhole protocol with $N$$=$$20$, $q$$=$$4$, $\beta$$=$$4$, ${\cal J}$$=$$1$, $t_0$$=$$2.4$,
$\mu$$=$$-0.2$, using $V^b$ to generate the
Left-Right interaction and as the size operator. 
The upper dashed black line is a linear fit to the Left size winding, weighted by the probability $P^\ell(s)$;
the lower dashed black line is the corresponding right size winding, and
the orange dashed line is a linear fit to the actual
points after applying the Left-Right interaction. 
The x-axis is the size as measured by $V^b$, as defined in
Eq. \ref{eq:size2op}.
{\bf Right:} The mutual information asymmetry for the same
parameters, fixing also $t_1$$=$$t_0$; the dashed line shows
$t_0$$=$$2.4$.
\label{fig:q4t24comp}}
\end{center}
\vspace{-0.5cm}
\end{figure}

Figure \ref{fig:q4t24comp} shows size winding using $V^b$ 
for the Left-Right
interaction and as the size operator, for the case
$N$$=$$20$, $q$$=$$4$, $\beta$$=$$4$, ${\cal J}$$=$$1$, $t_0$$=$$t_1$$=$$2.4$, $\mu$$=$$-0.2$.

\vspace{-0.3cm}

\begin{figure}[htb]
\vspace{-0.5 cm}
\begin{center}
\includegraphics[width=0.41\textwidth]{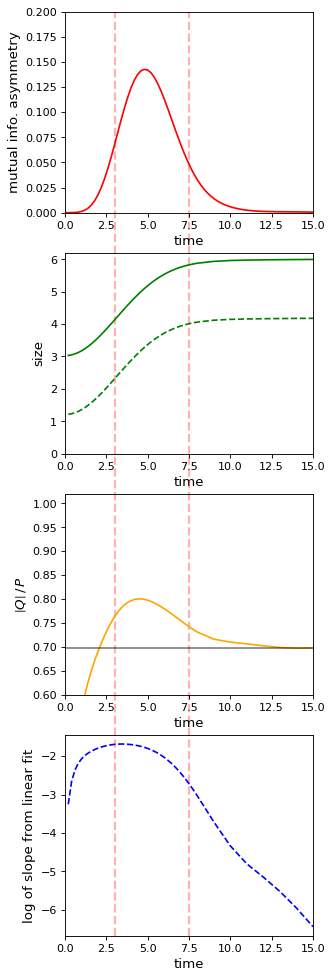}
\protect\caption{Size winding of the long-range wormhole
teleportation protocol with $N$$=$$24$, $q$$=$$8$, ${\cal J}$$=$$1$,
$\beta$$=$$8$, and $\mu$$=\pm 0.2$. {\bf Top:} The mutual information asymmetry as a function of time. {\bf Second:} The mean size of the
thermal fermion operator. {\bf Third:} $|Q|/P$ (orange line) averaged over size;
the solid black line denotes $G(\beta )$.
{\bf Bottom:} The log of the fitted linear size winding slope.
\label{fig:allVb}}
\end{center}
\end{figure}

\subsection{Extracting the Lyapunov exponent from size winding}
As already noted, operator size growth from scrambling in the SYK model is related to the Lyapunov exponent, but in our finite $N$ examples
it is not possible to extract this exponent directly from the
observed growth rate over the relevant time range.
However, as observed by \citet{nezami2021quantum}, in the
large $N$, large $q$ limit of wormhole teleportation,
as one approaches
the scrambling time, the magnitude of the size winding slope
decreases exponentially like exp$(-\lambda t)$, where
$\lambda$ is again the Lyapunov exponent.

We see a corresponding behavior in our finite $N$, finite $q$
examples. This is shown for the Gao-Jafferis
protocol in Figure \ref{fig:all}, for the
same cases as Figure \ref{fig:sizegrowth}: the first two
plots starting from the top are the same as in
Figure \ref{fig:sizegrowth}. The third plot shows the 
coherence diagnostic, i.e., the value of $|Q(s)|/P(s)$ averaged
over size; notice that the coherence peaks in the time interval
where the mutual information asymmetry peaks. The
bottom plot shows the log of the magnitude of the fitted linear size winding slope;
this also peaks in the time interval
where the mutual information asymmetry peaks.
Furthermore at later times the plot shows an exponential
decline over time ranging from about 7.5 to the scrambling time 15.
Figure \ref{fig:allVb} demonstrates similar behavior for the
long-range wormhole teleportation protocol.

\begin{figure}[htb]
\vspace{-0.5 cm}
\begin{center}
\includegraphics[width=0.32\textwidth]{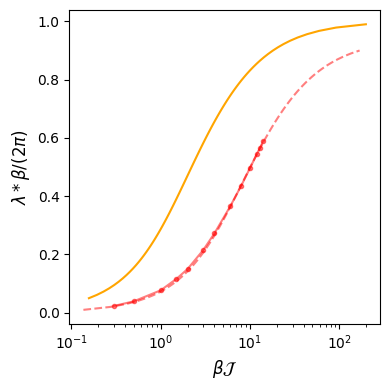}
\includegraphics[width=0.32\textwidth]{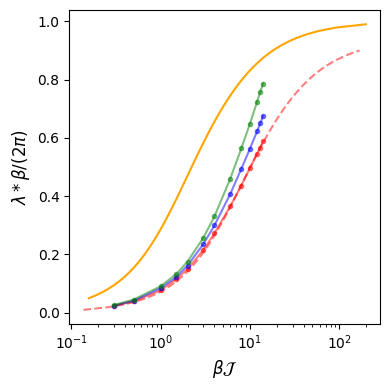}
\includegraphics[width=0.32\textwidth]{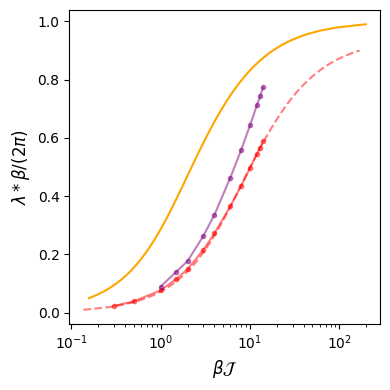}
\protect\caption{Lyapunov exponents. {\bf Left:} The orange line is the
SYK model Lyapunov exponent as a function of $\beta{\cal J}$,
calculated in the large $N$, large $q$ limit \cite{Maldacena:2016hyu}.
The red points are exponents extracted from the fitted size winding
slope of the SYK model with $N$$=$$26$, $q$$=$$8$, ${\cal J}$$=$$1$,
fitting to an exponential over times ranging from 7.5 to 15.
The dashed red line is a fit to the red points using a functional
form similar to the orange line.
{\bf Middle:} Lyapunov exponents for different $N$. 
The red points are exponents extracted from the fitted size winding
slope of the SYK model with $N$$=$$26$, $q$$=$$8$, ${\cal J}$$=$$1$,
fitting to an exponential over times ranging from 7.5 to 15.
Corresponding results are also shown for $N$$=$$24$ (blue points)
and $N$$=$$22$ (green points). {\bf Right:} The same as the left plot,
but adding the corresponding results for size winding
with sizes measured by $V^b$ (purple points).
\label{fig:Lyapunov}}
\end{center}
\end{figure}

Figure \ref{fig:Lyapunov} shows the Lyapunov exponents as a function
of $\beta$ as extracted from the size winding slope exponential dropoff.
Here we have used the largest values of $N$ and $q$ available
to us, attempting to make contact with the known analytic result
for the large $N$, large $q$ limit \cite{Maldacena:2016hyu}. 
In the plot we only use
values of beta small enough that $\lambda t \gtrapprox 2$ over
the fitted time range, to ensure that we are in the exponential
regime.

The qualitative agreement with the large $N$, large $q$ result
is encouraging, and resembles the large $N$, finite $q$
result obtained numerically and shown in Figure 11 of
\citet{Maldacena:2016hyu}. One would expect the agreement to
worsen as we decrease $N$ and indeed this is the case, as shown in 
Figure \ref{fig:Lyapunov}.

\begin{figure}[htb]
\vspace{-0.5 cm}
\begin{center}
\includegraphics[width=0.21\textwidth]{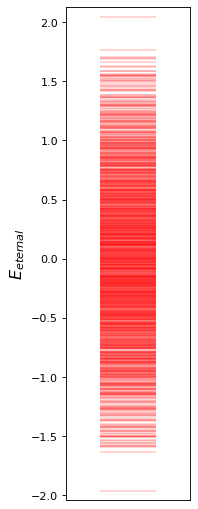}
\includegraphics[width=0.21\textwidth]{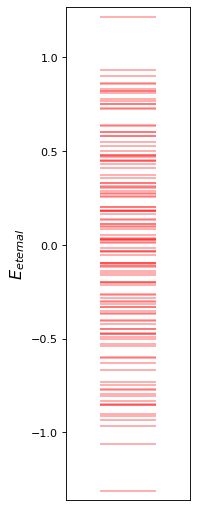}
\includegraphics[width=0.21\textwidth]{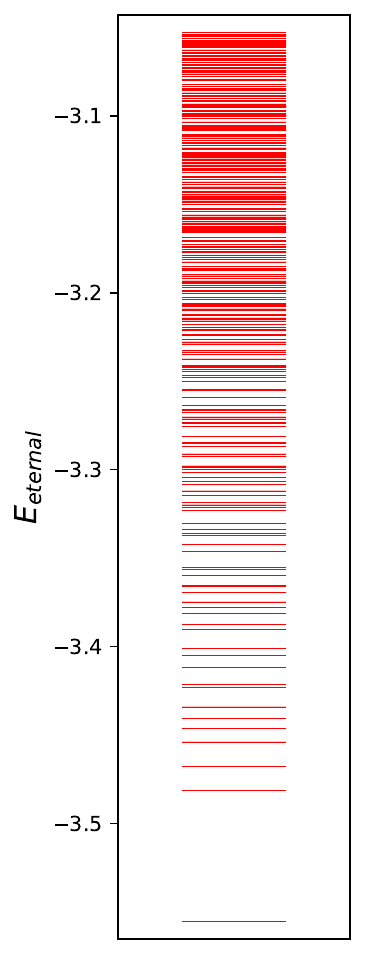}
\protect\caption{Spectra of the eternal traversable wormhole. {\bf Left:} Using the $N$$=$$10$,
$q$$=$$4$, ${\cal J}$$=$$1$ SYK model, averaged over 10 instantiations, and 
the interaction $V$ with $\mu$$=$$0.3$.
{\bf Middle:} Using the Learned17 model with $\mu$$=$$0.3$. 
{\bf Right} Using the $N$$=$$18$,
$q$$=$$4$, ${\cal J}$$=$$1$ SYK model, averaged over 10 instantiations, and 
the interaction $V^b$ with $\mu$$=$$0.3$; only the lowest 600 energy eigenstates are shown.
\label{fig:e_spectrum_N10}}
\end{center}
\end{figure}

\subsection{Eternal traversable wormhole}\label{sec:eternal}

As discussed by \citet{Maldacena:2018lmt}, The holographic dual 
description can be elucidated by studying properties of the
``eternal" traversable wormhole evident in the dynamics of the
Hamiltonian
\begin{align}
\label{eq:Heternal}
    H_{\rm eternal} = H_L + H_R + \mu V
\end{align}
where here positive values of $\mu$ correspond to negative values of $\mu$
in the wormhole teleportation protocol (this is just the sign difference between
applying a Left-Right interaction exp$(i\mu V)$ and performing time evolution
with exp$(-i(H_L + H_R + \mu V)t)$).

\begin{figure}[htb]
\vspace{-0.5cm}
\begin{center}
\includegraphics[width=0.45\textwidth]{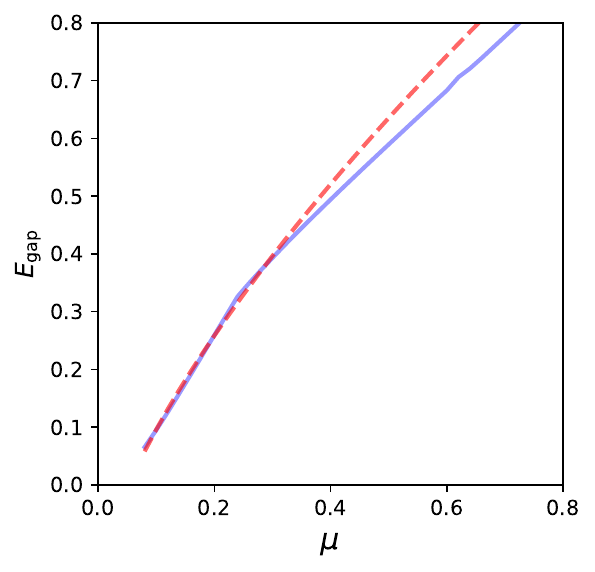}
\protect\caption{The energy gap of the eternal traversable wormhole
as a function of $\mu$, for $N$$=$$20$, $q$$=$$4$, ${\cal J}$$=$$1$
(blue solid line). The red dashed line shows a fit of the
$\mu < 0.3$ portion of this result to a simple power law
behavior $E_{\rm gap} = a\mu^b + c$, with $a,b,c$ = 1.3, 0.69, and
-0.17 respectively.
\label{fig:e_fit}}
\end{center}
\end{figure}

The approximate $SL(2,R)$ remnant of reparametrization invariance
is here related to a gap in the energy spectrum of $H_{\rm eternal}$.
This gap is a robust feature evident already for $N$$=$$10$, as seen
in Fig. \ref{fig:e_spectrum_N10}. Here we also show the spectrum
of our learned Hamiltonian used in the Google Sycamore experiment described in \cite{jafferis2022traversable}. We henceforth
refer to this as  {\it Learned17}, since for this
model $H_{\rm eternal}$ has 17 terms.  In the large $N$ limit,
\citet{Maldacena:2018lmt} showed that the dependence of the energy
gap on the coupling $\mu$ has two regimes: for large $\mu$ the
energy gap grows linearly with $\mu$, while for small $\mu$ the 
dependence is related to the holographic $SL(2,R)$ symmetry.
In the large $N$ limit with $q$$=$$4$, the prediction for the small $\mu$
regime is
$E_{\rm gap} \propto \mu^{2/3}$. Fig. \ref{fig:e_fit} shows the results of
our simulation for $N$$=$$20$; the two distinct regimes are clearly visible,
and a simple fit to the small $\mu$ behavior gives 
$E_{\rm gap} \propto \mu^{0.69}$, consistent with the large $N$ result.
Similar results at finite $N$ were obtained in \cite{Garcia-Garcia:2019poj}.

As detailed in \cite{Maldacena:2018lmt}, one expects the 
thermofield double state $\ket{\rm tfd}$ of the wormhole teleportation protocol
to approximate the ground state $\ket{\rm G}$ of $H_{\rm eternal}$.
Indeed the overlap $\bra{\rm tfd}\ket{\rm G}$ approaches unity in
two limits. The first limit is $\mu$ becoming large, 
where $\ket{\rm G} \to \ket{I}$, since
as we saw in Eq. \ref{eq:sizeop}, this maximally entangled state yields
the minimum expectation value of the size operator $V$.
Since $\ket{I}$ is also the infinite temperature limit of $\ket{\rm tfd}$,
we see that $\bra{\rm tfd}\ket{\rm G} \to 1$ as $\mu \to \infty$ and $\beta \to 0$. The second limit is $\mu \to 0$, $\beta \to \infty$, since in this limit
$H_{\rm eternal} \to H_L + H_R$, and large $\beta$ concentrates the support of
the $\ket{\rm tfd}$ onto the SYK ground state.

\begin{figure}[htb]
\vspace{-0.5cm}
\begin{center}
\includegraphics[width=0.4\textwidth]{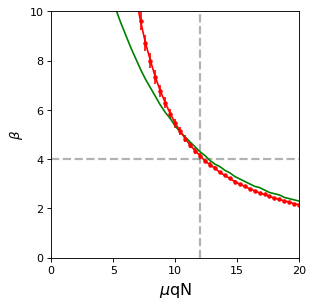}
\includegraphics[width=0.4\textwidth]{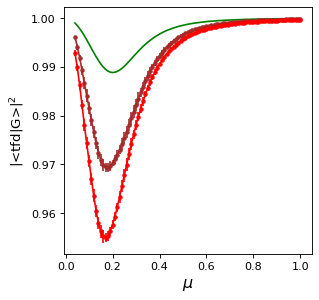}
\protect\caption{{\bf Left:} The relation between $\beta$ and $\mu$ obtained by
maximizing the overlap $\bra{\rm tfd}\ket{\rm G}$, shown for $H_{\rm eternal}$
with $N$$=$$10$, $q$$=$$4$, ${\cal J}$$=$$1$ averaged over 10 instantiations
(red points), as well as the Learned17 model (green line) of
\cite{jafferis2022traversable}. The dashed lines show the values
of $\mu$ (scaled by $qN$ as in \cite{jafferis2022traversable}), and $\beta$
that give the optimal mutual information transfer in the wormhole teleportation
protocol, as described in \cite{jafferis2022traversable}.
{\bf Right:}
The maximal overlap of the $H_{\rm eternal}$ ground state and the thermofield double state, as a function of $\mu$, with $H_L$,$H_R$
taken from the
$N$$=$$10$, $q$$=$$4$, ${\cal J}$$=$$1$ SYK model averaged over 10 instantiations (red points), the $N$$=$$8$, $q$$=$$4$, ${\cal J}$$=$$1$ SYK model averaged over 10 instantiations (brown points), 
and the Learned17 model of \cite{jafferis2022traversable} (green line).
\label{fig:beta_vs_mu}}
\end{center}
\end{figure}

In between these two limits, one can adjust $\beta$ as a function of $\mu$
to maximize the overlap $\bra{\rm tfd}\ket{\rm G}$. This leads to curves
like those in Fig. \ref{fig:beta_vs_mu}. It is interesting to note from
this figure that for the $N$$=$$10$, $q$$=$$4$, ${\cal J}$$=$$1$ model
first discussed in \cite{jafferis2022traversable}, optimizing the overlap
in $\bra{\rm tfd}\ket{\rm G}$ gives a relation between $\beta$ and $\mu$
consistent with that obtained by asking for optimal mutual information transfer in the wormhole teleportation protocol. Furthermore, the Learned17 model
of \cite{jafferis2022traversable} tracks the $\beta$ vs $\mu$ behavior of its parent SYK-based model over a large range. 

If we now fix the relation between $\beta$ and $\mu$ by the criterion
of maximizing the overlap of the $H_{\rm eternal}$ ground state and the
thermofield double state, we can plot the maximal overlap as a function of
$\mu$, as shown in Fig. \ref{fig:beta_vs_mu} with $H_L$,$H_R$ taken
from three example Hamiltonians: the
$N$$=$$10$, $q$$=$$4$, ${\cal J}$$=$$1$ SYK model, the $N$$=$$8$, $q$$=$$4$, ${\cal J}$$=$$1$ SYK model, 
and the Learned17 model of \cite{jafferis2022traversable}.
In all cases (as previously noted for several examples in \cite{Maldacena:2018lmt}) the
overlap is close to unity over the entire range.

In the wormhole teleportation protocol, the Left-Right interaction
exp$(i\mu V)$ is applied at $t$$=$$0$ when the state of the system, by construction,
is close to $\ket{\rm tfd}$. It is thus interesting to construct a measure
of how well $\ket{\rm tfd}$ respects the part of the approximate $SL(2,R)$
symmetry that generates global time translations. A reasonable
figure of merit is
\begin{align}
\label{eq:SL2R}
    \frac{\bra{\rm tfd} (H_{\rm eternal} - E_0) \ket{\rm tfd}}{|E_0|}
\end{align}
where $E_0$ is the energy of the true ground state $\ket{\rm G}$ of
$H_{\rm eternal}$. When this quantity is small, the 
expectation values of all three $SL(2,R)$ generators in the thermofield double state either vanish or are small in natural units.

\begin{figure}[htb]
\begin{center}
\includegraphics[width=0.41\textwidth]{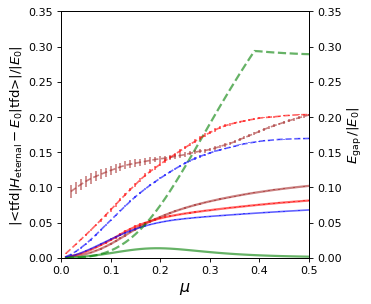}
\includegraphics[width=0.41\textwidth]{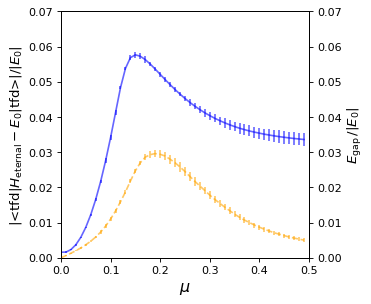}
\protect\caption{$SL(2,R)$ symmetry. {\bf Left:} The figure of merit defined in Eq. \ref{eq:SL2R}, 
as a function of $\mu$, for the Gao-Jafferis wormhole teleportation
protocol with $H_L$,$H_R$
taken from the $q$$=$$4$, ${\cal J}$$=$$1$ SYK model averaged over 10 instantiations with $N$$=$$12$ (blue solid), $N$$=$$10$ (red solid),  $N$$=$$8$ averaged over 40 instantiations (brown solid), 
and the Learned17 model of \cite{jafferis2022traversable} (green solid).
Shown on the same scale, with the same color coding but dashed lines, is the corresponding value of
$E_{\rm gap}$, also in units of $|E_0|$.
{\bf Right} The same figure of merit (blue points) for the long-range wormhole teleportation protocol,
with $H_L$,$H_R$
taken from the $N$$=$$18$, $q$$=$$4$, ${\cal J}$$=$$1$ SYK model averaged over 20 instantiations.
Shown on the same scale (orange dashed lines) is the corresponding value of
$E_{\rm gap}$ in units of $|E_0|$.
In all cases $\beta$ is set to the value that minimizes
$\bra{\rm tfd} (H_{\rm eternal}$$-$$E_0) \ket{\rm tfd}/|E_0|$. 
\label{fig:SL2R}}
\end{center}
\vspace{-0.4cm}
\end{figure}

We plot this quantity in Fig. \ref{fig:SL2R} (Left) for several examples
of the Gao-Jafferis protocol.
Recall that $\mu$$\simeq$$0.3$ optimizes
the mutual information transfer in the wormhole teleportation protocol
for both the $N$$=$$10$ and the Learned17 models, as discussed in
\cite{jafferis2022traversable}. In the figure we show for comparison
$E_{\rm gap}/|E_0|$, i.e., the energy gap in units of the magnitude
of the ground state energy.
Notice that in all four examples the figure of merit is both numerically
small and smaller than the rescaled gap energy;
this can be taken as an indication that the $SL(2,R)$
symmetry is indeed approximately respected.

\subsection{Modified Eternal Hamiltonian}

We perform a similar analysis now for a modified version of the Hamiltonian in Eq. \ref{eq:Heternal}, where we replace the
interaction operator $V$ with $V^b$:
\begin{align}
\label{eq:HeternalVb}
    H^b_{\rm eternal} = H_L + H_R + \mu V^b
\end{align}

It is useful at this point to identify some discrete symmetries
respected by these two eternal Hamiltonians.
As already discussed by
\citet{Garcia-Garcia:2019poj}, the original eternal
Hamiltonian of Eq. \ref{eq:Heternal} commutes with the operator $Q$ defined by
\begin{align}
\label{eq:Qopdef}
    Q \equiv {\rm e}^{i\pi V/2}
\end{align}
The operator $Q$ has eigenvalues: $\pm 1$, $\pm i$.
The square of $Q$ is the overall parity operator $\Gamma^5$:
\begin{align}
\label{eq:paritydef}
    Q^2 = \Gamma^5 = \Gamma^5_R \cdot \Gamma^5_L
\; ;\quad{\rm where:}\quad
    \Gamma^5_L = \prod_{j=0}^{N-1}\Gamma^{(1)\ell}_j \; ;\quad
    \Gamma^5_R = \prod_{j=0}^{N-1}\Gamma^{(1)r}_j
\end{align}
The Left and Right side parity operators $\Gamma^5_L$ and $\Gamma^5_R$
commute with each other and have eigenvalues $q_L,q_R$$=$$\pm 1$, but
they do not commute with $H_{\rm eternal}$.
Obviously $\Gamma^5$ does commutes with the eternal Hamiltonian and 
also has eigenvalues $\pm 1$.

\begin{figure}[htb]
\vspace{-0.5cm}
\begin{center}
\includegraphics[width=0.41\textwidth]{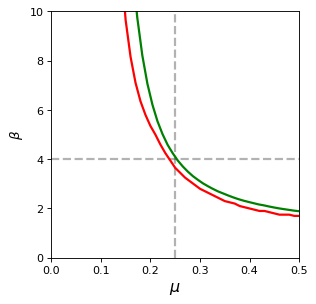}
\includegraphics[width=0.41\textwidth]{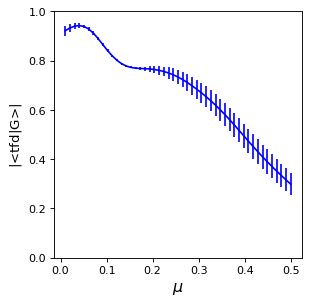}
\protect\caption{{\bf Left:} The relation between $\beta$ and $\mu$ obtained by
maximizing the overlap $\bra{\rm tfd}\ket{\rm G}$, shown for 
$H_{\rm eternal}$$=$$H_L$$+$$H_R$$+$$\mu V^b$
with $q$$=$$4$, ${\cal J}$$=$$1$ and $N$$=$$18$ averaged over 20 instantiations
(green points), as well as $N$$=$$20$ averaged over 10 instantiations
(red points),
The dashed lines show the approximate values
of $\mu$ and $\beta$
that give the optimal mutual information transfer in the 
long-range wormhole teleportation
protocol.
{\bf Right:}
The maximal overlap of the modified $H_{\rm eternal}$ ground state and the thermofield double state, as a function of $\mu$, with $H_L$,$H_R$
taken from the
$N$$=$$18$, $q$$=$$4$, ${\cal J}$$=$$1$ SYK model averaged over 20
instantiations.
\label{fig:beta_vs_mu_Vb}}
\end{center}
\end{figure}

Our modified eternal Hamiltonian $H^b_{\rm eternal}$ has 
additional discrete symmetries. It commutes with $Q$, and also commutes
with the Left/Right parity operators $\Gamma^5_L$, and $\Gamma^5_R$ individually, in addition to the total parity operator $\Gamma^5$.

From Eqs. \ref{eq:Qopdef} and \ref{eq:paritydef} one can easily show
that the discrete symmetry operators $Q$ and $\Gamma^5_L$ commute with
each other when $N$$=$$0$ (mod 4), but anti-commute with each other
when $N$$=$$2$ (mod 4).

Figure \ref{fig:e_spectrum_N10} (Right) shows the 600 lowest energy
eigenvalues of $H^b_{\rm eternal}$ using the
$N$$=$$18$,
$q$$=$$4$, ${\cal J}$$=$$1$ SYK model, averaged over 10 instantiations, and 
the interaction $V^b$ with $\mu$$=$$0.3$. 
Since $N$$=$$18$$=$$2$ (mod 4), $Q$ acting on an eigenstate of $\Gamma^5_L$, 
or vice versa, gives a new state with the same energy eigenvalue.
Thus in this spectrum all of the energy eigenstates are doubly degenerate.
The ground state pair has $q_Q$$=$$\pm i$ with a suitable choice of basis,
and has $q_L$$=$$\pm 1$ in a different suitable choice of basis. We see in the
figure that an
energy gap is evident; more generally we find that
the spectra of $H^b_{\rm eternal}$ and $H_{\rm eternal}$ are qualitatively similar. 

Figure \ref{fig:beta_vs_mu_Vb} (Left) plots the value of 
$\beta$ as a function of $\mu$ that maximizes
the overlap $\bra{\rm tfd}\ket{\rm G}$, where $\ket{\rm G}$ now denotes
the ground state of the modified eternal Hamiltonian Eq. \ref{eq:HeternalVb}.
As for the original eternal Hamiltonian Eq. \ref{eq:Heternal}, 
optimizing this overlap gives a relation between $\beta$ and $\mu$
consistent with that obtained by asking for optimal mutual information transfer in the wormhole teleportation protocol. 
In Figure \ref{fig:beta_vs_mu_Vb} (Right) we see that the overlap
$|\bra{\rm tfd}\ket{\rm G}|$ is fairly large in the relevant
range of $\mu$, but significantly smaller than
we had found in Figure \ref{fig:beta_vs_mu} for the original
eternal Hamiltonian.

Figure \ref{fig:SL2R} (Right) shows the figure of merit defined in Eq. \ref{eq:SL2R} for the long-range wormhole teleportation protocol,
as a function of $\mu$ for the modified
eternal Hamiltonian $H^b_{\rm eternal}$
with $N$$=$$18$, $q$$=$$4$, ${\cal J}$$=$$1$.
Shown on the same scale is the corresponding value of the energy gap
in units of $|E_0|$. Comparing to the results for 
$H_{\rm eternal}$ shown in Figure \ref{fig:SL2R} (Left),
note that for $H^b_{\rm eternal}$ the figure of merit is numerically quite small, but not smaller than the rescaled energy gap.

\begin{figure}[htb]
\begin{center}
\includegraphics[width=0.45\textwidth]{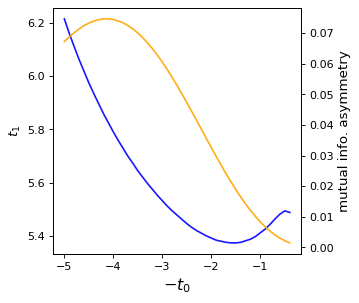}
\includegraphics[width=0.45\textwidth]{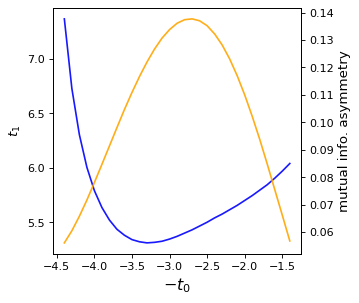}
\protect\caption{Causal time ordering of long-range wormhole
teleportation.
{\bf Left:}
$N$$=$$24$, $q$$=$$8$, ${\cal J}$$=$$1$,
$\beta$$=$$20$. The Left-Right interaction using $V^b$ is applied
at time $t$$=$$0$, with coupling strength $\mu$$=$$-0.18$.
The blue line shows the extraction time $t_1$
as a function of insertion time $t_0$, as determined by maximizing
the resulting mutual information asymmetry. The orange line shows
the corresponding mutual information asymmetry.
{\bf Right:}
$N$$=$$26$, $q$$=$$8$, ${\cal J}$$=$$1$,
$\beta$$=$$16$. The Left-Right interaction using $V^b$ is applied
twice, with coupling strength $\mu$$=$$-0.11$, at times
$t$$=$$\pm 1.5$. 
\label{fig:timeordering}}
\end{center}
\vspace{-0.1cm}
\end{figure}

\section{Causal time ordering}\label{sec:timeordering}

Figure \ref{fig:timeordering} illustrates two examples of the
causal time ordering of long-range wormhole
teleportation. For large 
$t_0$,$t_1$ (left side of each plot), the ``peaked-size teleporation''
dominates, and the time ordering is ``first in, last out", as
one would expect for any many-body mechanism based on scrambling
followed by unscrambling.
For times when the holographic wormhole teleportation dominates
(right side of each plot), we see instead the causal
time ordering ``first in, first out", that one would expect from
traversing through a wormhole.

The example in the left plot uses
$N$$=$$24$, $q$$=$$8$, ${\cal J}$$=$$1$,
$\beta$$=$$20$, and the Left-Right interaction using $V^b$ is applied
at time $t$$=$$0$, with coupling strength $\mu$$=$$-0.18$.
The blue line shows the extraction time $t_1$
as a function of insertion time $t_0$, as determined by maximizing
the resulting mutual information asymmetry. The orange line shows
the corresponding mutual information asymmetry.
While the causal time ordering is evident at early times, at later
times the opposite time ordering, which arises from ``peaked-size teleportation'' and has steeper intrinsic slope, quickly dominates.
Since some amount of ``peaked-size teleportation'' is always present
during the period when the holographic wormhole teleportation is
operative, it is non-trivial to predict over what time interval, if any,
the causal time ordering dominates.

The example in the right plot uses
$N$$=$$26$, $q$$=$$8$, ${\cal J}$$=$$1$,
$\beta$$=$$16$. In this example, instead of applying
the Left-Right interaction using $V^b$
at $t$$=$$0$, with coupling strength $\mu$$=$$-0.18$, we
instead apply the Left-Right interaction
twice, with a weaker coupling strength $\mu$$=$$-0.11$, at times
$t$$=$$\pm 1.5$. Applying the Left-Right interaction in multiple
time slices disfavors the ``peaked-size teleportation'' 
in favor of the holographic wormhole mechanism, increasing the time interval over which the causal time ordering is apparent.
This is a general feature that applies also to the original
Gao-Jafferis protocol, as noted in \cite{jafferis2022traversable}.

\section{Measurement, multipartite entanglement, and a classical channel}\label{sec:classical}

In order to implement a classical channel for
long-range wormhole teleportation, two modifications of the
protocol are required:
\begin{itemize}
\item At time $t=0$, instead of applying the Left-Right
interaction operator exp$(i\mu V^b)$, Alice measures the
$2^{N/2}$ Left side qubits and communicates the result to Bob through
a classical channel.
\item Bob applies the Right side operator 
\begin{align}
\label{eq:Ropdef}
    V^{cc} \equiv {\rm exp}\left(i\mu \sum_{j=0}^{N/2-1} 
    \, s_j\, Z_{j+N/2} \right)
\end{align}
where $s_j$ are the Pauli $Z$ eigenvalues $\pm1$ of the
measured left qubits.
\item At time $t=t_1$, Bob swaps the first right qubit
with the qubit $T$.
\item If the measured parity $\Gamma^5_\ell$ was
-1, Bob also applies a Pauli $Z$ gate to the qubit $T$.
\end{itemize}

To understand the role of multipartite entanglement
in this protocol, we start with the simple
case of the protocol with $t_0$$=$$t_1$$=$$\beta$$=$0.
Since in this case there is no scrambling, we only need
to consider the 5-qubit system $RQ\ell r T$, where 
$\ell$,$r$ refer to the first Left and first Right qubits,
respectively. The $\beta$$=$$0$ initial state is
\begin{align}
\label{eq:instate}
    \frac{1}{\sqrt{2}}\left(
    \ket{00}_{RQ} + \ket{11}_{RQ} \right)
    \otimes \frac{1}{\sqrt{2}}\left(
    \ket{01}_{\ell r} + i\ket{10}_{\ell r} \right)
    \otimes \ket{0}
\end{align}
After applying the long-range protocol, the final state of
the five qubits $RQ\ell rT$ is:
\begin{align}
\label{eq:outstate}
    \frac{1}{2}\left( {\rm e}^{-i\mu}\ket{00001}
    +i{\rm e}^{i\mu}\ket{01000}
    -{\rm e}^{i\mu}\ket{10101}
    +i{\rm e}^{-i\mu}\ket{11100} \right)
\end{align}
For the moment we consider the quantum channel with
no measurement of the Left qubit $\ell$. Then it is easy to show that
\begin{align}
\label{eq:Iresults}
    I(R:T) = 0\; ;\quad
    I(R:\ell) = {\rm log}_2\left(3 - {\rm cos}(4\mu) \right)
    \; ;\quad I(R:\ell T) = 2 \\
    {\rm tr}\left( \rho_{R\ell}^2 \right) = \frac{1}{2} 
    + 2\,{\rm cos}^2\mu \,{\rm sin}^2\mu 
    \qquad\qquad\qquad\qquad\qquad\qquad\qquad\quad
    \nonumber
\end{align}
We see that there is no quantum information shared
bilaterally between qubits $R$ and $T$, as expected.
However for almost all values of the coupling $\mu$
there is tripartite mutual information 
$I(R$$:$$\ell$$:$$T)$ shared between
qubits $R$, $\ell$, and $T$. This is computed as
\begin{align}
\label{eq:tripartite}
    I(R:\ell :T) =
    I(R:T) + I(R:\ell ) - I(R:\ell T)
    = -2 + {\rm log}_2\left(3 - {\rm cos}(4\mu) \right)
\end{align}
Thus, for example, for $\mu$$=$0 we get
$I(R$$:$$T)$$=$$0$, $I(R$$:$$\ell)$$=$$1$, $I(R$$:$$\ell T)$$=$$2$,
and thus
$I(R$$:$$\ell$$:$$T)$$=$$-1$, the minimum (i.e. most negative)
possible value. In this example the 3-qubit system 
$R\ell T$ in the final state is sharing one qubit worth
of quantum information, with part of it
shared bilaterally between $R$ and $\ell$,
no information shared
bilaterally between $R$ and $T$, and the rest
shared non-locally.

\begin{figure}[htb]
\begin{center}
\includegraphics[width=0.42\textwidth]{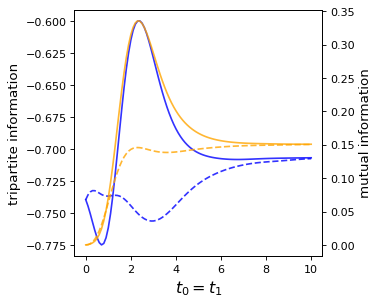}
\includegraphics[width=0.42\textwidth]{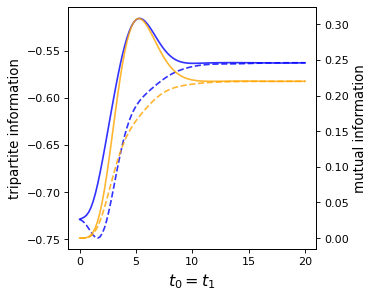}
\protect\caption{Multipartite quantum information sharing
in the long-range wormhole teleportation protocol. {\bf Left:} Using 
$N$$=$$24$, $q$$=$$4$, ${\cal J}$$=$$1$,
$\beta$$=$$4$, and $|\mu|$$=$0.2, for a single instantiation
of the SYK Hamiltonian.
The blue solid/dashed lines show the tripartite information
$I(R$$:$$L$$:$$T)$, where $L$ refers collectively 
to the 12 Left side qubits,
as a function of $t_0$$=$$t_1$, for $\mu$ negative/positive.
For comparison, the orange solid/dashed lines show the 
mutual information $I(R$$:$$T)$ for $\mu$ negative/positive.
{\bf Right:} Same thing for $N$$=$$24$, $q$$=$$8$, ${\cal J}$$=$$1$,
$\beta$$=$$8$, and $|\mu|$$=$0.2,
\label{fig:tripartite}}
\end{center}
\end{figure}

Mutltipartite quantum information is a characteristic
feature of scrambled systems \cite{Hosur:2015ylk}, as we will see, but in this simple
example there was no scrambling. Similarly holographic
systems are known \cite{Hayden:2011ag} generally to exhibit negative (or vanishing) tripartite information,i.e. they are ``superextensive", 
but in this simple example holography is not relevant.
It is also worth observing, as noted in \cite{Hosur:2015ylk}, 
that tripartite
quantum information should really be regarded as a possible
result of 4-body entanglement; in our example this is the
statement that the initial state entanglement of qubits
$\ell$ and $r$ is important.

To conclude our simple example consider the case
$\mu$$=$$\pi/4$; now we get
$I(R$$:$$\ell )$$=$$2$ and 
${\rm tr}( \rho_{R\ell}^2 )$$=$$1$,
so the qubits $R$ and $\ell$ are sharing all of their
quantum information bilaterally, and
the tripartite information $I(R$$:$$\ell$$:$$T)$ vanishes.

Figure \ref{fig:tripartite} shows 
$I(R$$:$$L$$:$$T)$, where $L$ refers collectively to all the 
Left side qubits,
as a function of $t_0$$=$$t_1$, for
the long-range wormhole teleportation protocol with 
$N$$=$$24$, ${\cal J}$$=$$1$, $|\mu|$$=$$0.2$, and either
$q$$=$$4$, $\beta$$=$$4$ (Left) or $q$$=$$8$, $\beta$$=$$8$ (Right).
We see that $I(R$$:$$L$$:$$T)$ is negative for all values
of $t_0$, as expected. 
The tripartite information as a function of $t_0$$=$$t_1$ tracks
the mutual information $I(R$$:$$T)$ rather closely, and in the
first example has
an even more pronounced asymmetry between negative
and positive values of $\mu$.

\begin{figure}[htb]
\begin{center}
\includegraphics[width=0.42\textwidth]{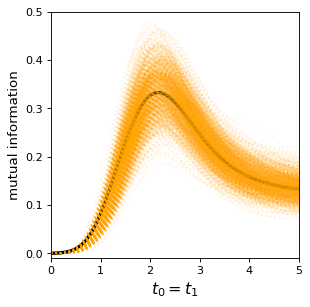}
\includegraphics[width=0.42\textwidth]{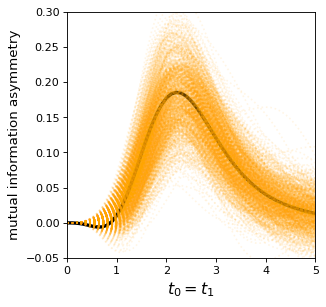}
\protect\caption{The effect of measuring the Left side qubits
upon the end state mutual information of qubits $R$ and $T$
in the long-range wormhole teleportation protocol.
The figures show the mutual information as a function of $t_0$$=$$t_1$
and the mutual information asymmetry
between negative/positive values of $\mu$, for 
$N$$=$$20$, $q$$=$$4$, ${\cal J}$$=$$1$,
$\beta$$=$$4$, and $|\mu|$$=$0.25, for a single instantiation
of the SYK Hamiltonian. The black line in each figure denotes the result
without measurement, while the orange dotted lines show
results for all of the 1024 possible measurement outcomes.
\label{fig:measure}}
\end{center}
\end{figure}

The fact that the qubits $R$ and $T$ 
generally have multipartite
entanglement with the Left side SYK qubits has a significant
effect on the results of the classical channel long-range
protocol. When the Left qubits are measured, this
multipartite entanglement collapses down to bilateral
entanglement between $R$ and $T$. The final value
of $I(R$$:$$T)$ in general depends on the measurement
outcome of the Left side qubits, and may be either larger or smaller
than the value in the quantum channel protocol with no
measurement.

Figure \ref{fig:measure} shows 
the effect of measuring the Left side qubits
upon the final state mutual information of qubits $R$ and $T$
in the long-range wormhole teleportation protocol.
Here we compare, for the case 
$N$$=$$20$, $q$$=$$4$, ${\cal J}$$=$$1$,
$\beta$$=$$4$, $|\mu|$$=$0.25, the
mutual information $I(R$$:$$T)$ using either no
measurement of the Left side qubits, or considering all
1024 possible measurement outcomes. We see that the
value of the mutual information asymmetry does indeed
depend significantly on the measurement outcome,
and may be either larger or smaller.

The holographic interpretation of the traversable wormhole
in the case of long-range wormhole teleportation with a classical
channel was already discussed in \citet{nezami2021quantum}, and
can be extracted from the related work of
\citet{Kourkoulou:2017zaj} and \citet{Antonini:2022lmg}.

\section{Outlook}\label{sec:outlook}

We demonstrate  a viable finite-$N$ long-range wormhole teleportation
protocol, using a classical channel, and exhibit its main properties via classical computer simulation. 
Here we discuss the possibility of performing long-range
wormhole teleportation on quantum hardware. 

A quantum hardware implementation of the
Gao-Jafferis protocol was reported in \citet{jafferis2022traversable}.
That experiment relied on two key enabling features: 
a performant quantum processor (Google Sycamore) and the use of the
Learned17 model. The latter was developed starting from a model
based on a single instantiation of $N$$=$$10$, $q$$=$$4$
SYK, with 430 operators total to describe $H_L$, $H_R$, and $V$. The learning procedure used gradient descent
with L1 regularization and a loss function attempting to preserve
a mutual information curve similar to Figure \ref{fig:self-average},
and produced a learned model with 5$+$5$+$7$=$17 operators total to describe $H_L$, $H_R$, and $V$. We checked that,
although it does not have single-sided late time 
thermalization \cite{Kobrin:2023rzr},
the learned model exhibits the
holographic features of wormhole teleportation, including size winding,
causal time ordering, and the approximate SL(2,R) symmetry, as shown
here and in \citet{jafferis2022traversable,Jafferis:2023moh}.
These are non-trivial checks since the gradient descent loss function
makes no reference to any of these features. Note also that the learning
procedure is completely distinct from other procedures introduced in the
literature to produce simplified versions of the SYK model, such as
random sparsification \cite{Xu:2020shn} and commuting models (see Appendix 
\ref{sec:commuting} and \cite{Gao:2023gta}).

Because of the difference between $V$ and $V^b$ as size operators
(compare Eqs. \ref{eq:sizeop} and \ref{eq:size2op}) we have seen that
in some respects $N$$=$$10$ Gao-Jafferis wormhole teleportation is similar to
$N$$=$$20$ long-range wormhole teleportation. Thus learning or some other
sparsification procedure will certainly be required to implement
the long-range protocol even in the quantum channel.

Even more challenging is the prospect of performing long-range wormhole
teleportation with the classical channel, thus transferring quantum
information between two entangled systems that are, say, 100 km apart.
Quantum teleportation systems at the 100 km scale over optical fiber
already exist for the conventional Alice-Bob protocol \cite{Valivarthi:2020yax}.
Long-range wormhole teleportation will require the ability, in addition, to prepare
highly entangled states of many qubits in the optical domain. 
Performing the time evolution scrambling will require suitable
quantum processors, either digital or 
analog \cite{Shackleton:2023lpw,Uhrich:2023ddx}.

\begin{acknowledgements}
We thank Adam Brown, Ping Gao, David Gross, Alexei Kitaev, Juan Maldacena, Sepehr Nezami, 
Subir Sachdev and Leonard Susskind for their
helpful comments and questions.
This work is supported by the Department of Energy Office of High Energy Physics QuantISED program grant SC0019219 on Quantum Communication Channels for Fundamental Physics. This manuscript has been authored by Fermi Research Alliance LLC under Contract No. DE-AC02-07CH11359 with the U.S. Department of Energy, Office of Science, Office of High Energy Physics. SID is partially supported by the Brinson Foundation.
\end{acknowledgements}

\bibliography{scibib}

\appendix

\section{Computation of the {\boldmath $t_0$$=$$t_1$$=$$\beta$$=$$0$} reduced density
matrix for RT}
The computation of the $t$$=$$0$, $\beta$$=$$0$ reduced density
matrix for the reference and readout qubits $RT$ is straightforward
using the following identities:
\begin{align}
\label{eq:identities}
\chi_\ell^\dagger\chi_\ell &= \frac{1}{2} (1 + \Gamma_0^{(2)\ell} ) \; ; \quad
\chi_\ell\chi_\ell^\dagger = \frac{1}{2} (1 - \Gamma_0^{(2)\ell} ) \nonumber\\
\chi_r^\dagger\chi_r &= \frac{1}{2} (1 + \Gamma_0^{(2)r} ) \; ; \quad
\chi_r\chi_r^\dagger = \frac{1}{2} (1 - \Gamma_0^{(2)r} ) \nonumber\\
\Gamma_0^{(2)r}|I\rangle &= -\Gamma_0^{(2)\ell}|I\rangle \; ;\quad\quad\,
\Gamma_0^{(2)r}\Gamma_0^{(2)\ell} |I\rangle = -|I\rangle \nonumber\\
\Gamma_0^{(2)r}\chi_\ell |I\rangle &= -\chi_\ell |I\rangle \; ;\quad\quad\;
\Gamma_0^{(2)r}\chi_\ell^\dagger |I\rangle = \chi_\ell^\dagger |I\rangle \\
\psi_0^r \Gamma_0^{(2)\ell} |I\rangle &= \psi_1^\ell |I\rangle \; ;\quad\quad\quad
\psi_1^r \Gamma_0^{(2)\ell} |I\rangle = -\psi_0^\ell |I\rangle \nonumber\\
\Gamma_0^{(2)r} \psi_0^\ell |I\rangle &= -i\psi_1^\ell |I\rangle \; ;\quad\quad
\Gamma_0^{(2)r} \psi_1^\ell |I\rangle = i\psi_0^\ell |I\rangle \nonumber
\end{align}
We also need to introduce the notation:
\begin{align}
V_{01} = i\left( \psi_0^\ell\psi_0^r + \psi_1^\ell\psi_1^r \right)
\; ;\quad
V_{0} = i\left( \psi_0^\ell\psi_0^r \right) \; ;\quad
V_{1} = i\left( \psi_1^\ell\psi_1^r \right) \nonumber
\end{align}
and use the identities:
\begin{align}
\label{eq:Videntitiesa}
\{ V_{01}, \Gamma_0^{(2)\ell} \} &= 0\; ;\quad\quad\quad\quad 
\{ V_{01}, \Gamma_0^{(2)r} \} = 0 \nonumber\\
V_{01} \chi_\ell |I\rangle &= 0 \; ;\quad\quad\quad\quad
V_{01} \chi_\ell^\dagger |I\rangle = 0 \nonumber\\
V_0 \chi_\ell &= -\chi_\ell^\dagger V_0 \; ;\quad\quad 
V_0 \chi_\ell^\dagger = -\chi_\ell V_0 \\
V_1 \chi_\ell &= \chi_\ell^\dagger V_0 \; ;\quad\quad\quad 
V_1 \chi_\ell^\dagger = \chi_\ell V_0 \nonumber
\end{align}
which imply:
\begin{align}
\label{eq:Videntitiesb}
{\rm e}^{-i\mu V_{01}} \Gamma_0^{(2)r} \,{\rm e}^{i\mu V_{01}} &= 
{\rm e}^{-2i\mu V_{01}} \Gamma_0^{(2)r} \; ;\quad
{\rm e}^{-2i\mu V_{01}} \Gamma_0^{(2)\ell} = \Gamma_0^{(2)\ell} \,{\rm e}^{2i\mu V_{01}} \nonumber\\
{\rm e}^{-2i\mu V_{01}} |I\rangle &= {\rm e}^{2i\mu} |I\rangle \; ;\quad
{\rm e}^{2i\mu V_{01}} |I\rangle = {\rm e}^{-2i\mu} |I\rangle \\
{\rm e}^{-2i\mu V_0}\chi_\ell |I\rangle &= -i\,{\rm sin}\mu |I\rangle \; ;\quad
{\rm e}^{-2i\mu V_1}\chi_\ell |I\rangle = -i\,{\rm sin}\mu |I\rangle \nonumber
\end{align}
Thus for example:
\begin{align}
\label{eq:rhoexamplea}
\rho_{11} &= \left\langle \chi_\ell\chi_\ell^\dagger 
\,{\rm e}^{-i\mu V} \chi_r\chi_r^\dagger \,{\rm e}^{i\mu V} 
\chi_\ell\chi_\ell^\dagger \right\rangle +
\left\langle \chi_\ell^\dagger 
\,{\rm e}^{-i\mu V} \chi_r\chi_r^\dagger \,{\rm e}^{i\mu V}
\chi_\ell \right\rangle \nonumber\\
&= \left\langle \frac{1}{2}(1-\Gamma_0^{(2)\ell} )\frac{1}{2}
\left( 1 - {\rm e}^{-2i\mu V_{01}} \Gamma_0^{(2)r} \right)
\frac{1}{2}(1-\Gamma_0^{(2)\ell} ) \right\rangle +
\left\langle \chi_\ell^\dagger \frac{1}{2} 
\left( 1 - {\rm e}^{-2i\mu V_{01}} \Gamma_0^{(2)r} \right)
\chi_\ell \right\rangle \nonumber\\
&= \frac{1}{8} \left\langle (1-\Gamma_0^{(2)\ell} )
\left( 1 - {\rm e}^{-2i\mu V_{01}} \right) (1-\Gamma_0^{(2)\ell} ) \right\rangle +
\frac{1}{2}\left\langle \chi_\ell^\dagger 
\left( 1 - {\rm e}^{-2i\mu V_{01}} \right)  \chi_\ell \right\rangle \nonumber\\
&= \frac{1}{8}\left\langle \left( 1 - {\rm e}^{-2i\mu V_{01}}  \right)
\right\rangle + 
\frac{1}{8}\left\langle \Gamma_0^{(2)\ell} 
\left( 1 - {\rm e}^{-2i\mu V_{01}}  \right)
\Gamma_0^{(2)\ell} \right\rangle +
\left\langle \chi_\ell^\dagger \chi_\ell \right\rangle \\
&= \frac{1}{8} \left( 1 - {\rm e}^{2i\mu} + 1 - {\rm e}^{-2i\mu} \right) + \frac{1}{2}
\nonumber\\
&= \frac{1}{2}(1+{\rm sin}^2\mu )\nonumber
\end{align}
another example:
\begin{align}
\label{eq:rhoexampleb}
\rho_{14} &= \left\langle \chi_\ell 
\,{\rm e}^{-i\mu V} \chi_r^\dagger \,{\rm e}^{i\mu V}
\chi_\ell\chi_\ell^\dagger \right\rangle +
\left\langle \chi_\ell^\dagger\chi_\ell 
\,{\rm e}^{-i\mu V} \chi_r^\dagger \,{\rm e}^{i\mu V}
\chi_\ell \right\rangle \nonumber\\
&= \frac{1}{4} \left\langle (\psi_0^\ell + i\psi_i^\ell )
\left( {\rm e}^{-2i\mu V_0}\psi_0^r - i\,{\rm e}^{-21\mu V_1}\psi_1^r \right)
(1-\Gamma_0^{(2)\ell} ) \right\rangle \nonumber\\
 &+ \frac{1}{4} \left\langle (1+ \Gamma_0^{(2)\ell} )
\left( {\rm e}^{-2i\mu V_0}\psi_0^r - i\,{\rm e}^{-21\mu V_1}\psi_1^r \right)
(\psi_0^\ell + i\psi_i^\ell ) \right\rangle \\
&= \frac{i}{2} \left\langle \chi_\ell 
\left( {\rm e}^{-2i\mu V_0} - {\rm e}^{-2i\mu V_1} \right)
\chi_\ell \right\rangle 
-\frac{i}{8} \left\langle (1+\Gamma_0^{(2)\ell} ) 
\left( {\rm e}^{-2i\mu V_0} + {\rm e}^{-2i\mu V_1} \right)
(1-\Gamma_0^{(2)\ell} ) \right\rangle \nonumber\\
&= \frac{i}{2}\left\langle \chi_\ell\chi_\ell^\dagger 
\left( i\,{\rm sin}(2\mu V_0) + i\,{\rm sin}(2\mu V_1) \right) 
\right\rangle \nonumber\\
&\quad -\frac{i}{8} \left\langle 
\left( {\rm e}^{-2i\mu V_0} + {\rm e}^{-2i\mu V_1} \right)
\right\rangle -\frac{i}{8} \left\langle \Gamma_0^{(2)\ell}
\left( {\rm e}^{-2i\mu V_0} + {\rm e}^{-2i\mu V_1} \right)
\Gamma_0^{(2)\ell} \right\rangle \nonumber\\
&= {\rm sin}\mu \nonumber
\end{align}

The final result for the  $t$$=$$0$, $\beta$$=$$0$ reduced density
matrix is shown is eqn. \ref{eq:rhoRTfinal}.

\section{Comparison with Gao's commuting models}\label{sec:commuting}
Here we make some comparisons with the non-holographic
commuting models recently introduced and studied by Ping Gao \cite{Gao:2023gta}.
The purpose is to exhibit to what extent these models can be distinguished
in terms of the properties of Gao-Jafferis teleportation.
For simplicity we will refer to the models of 
\cite{Gao:2023gta} as ``PG commuting 
models", and the models of \cite{jafferis2022traversable} as 
``SYK-based models". 

Gao-Jafferis wormhole teleportation is based 
on SYK couplings ${\cal J}_{ijkl}$ drawn from a random Gaussian ensemble
with mean zero and variance as given in Eq. \ref{eq:variance}.
For $q$$=$$4$ there are
${\cal O}(N^3)$ terms in the SYK Hamiltonian, while we scale the variance
by ${\cal O}(N^{-3})$. 

The PG commuting models of \cite{Gao:2023gta} are defined from a Hamiltonian
\begin{align}
\label{eq:PGham}
H_L = \sum_{(i,j)}  {\cal J}_{ij} 
X^\ell_i X^\ell_j
\end{align}
where 
\begin{align}
\label{eq:PGX}
X^\ell_i = \psi^\ell_{2i-2}\psi^\ell_{2i-1} \quad ; \quad i=1,\ldots,N/2
\end{align}
and the sum in the Hamiltonian is over all 
the distinct tuples
from choosing 2 distinct $X_i$ operators from $N/2$ possibilities.
The couplings ${\cal J}_{ij}$ are drawn from a random Gaussian ensemble
with mean zero and variance taken to be
\begin{align}
\label{eq:commuting-variance}
\langle ({\cal J}_{ij} )^2 \rangle = 
\frac{2^{q-1}(q/2-1)!(N/2 - q/2)!}{(N/2 - 1)!}{\cal J}^2
\end{align}
with ${\cal J}$ an overall fixed constant. Note that for $q=4$ there are
only ${\cal O}(N)$ terms in the commuting Hamiltonian, while we scale the variance by ${\cal O}(N^{-1})$. 

Taking $\cal{J}$ =1 in both cases, the variance of the ensemble used
for the PG commuting models is ${\cal O}(N^2)$ larger than the variance
of the ensemble used to define the corresponding SYK Hamiltonian.
This very large ratio of variances, numerically
$\gtrsim 100$ in the examples we consider, means that single instantiations
of the PG commuting models are never even a rough approximation to the
ensemble averages, and there is no sign of self-averaging in the PG commuting
models even taking the largest $N$ values, $N \sim 20$$-$$30$, for which we can perform the simulation of the corresponding SYK-based models.

\subsection{2-pt correlator of PG commuting models}

In \cite{Gao:2023gta} the Euclidean 2-pt correlator for Left majoranas of the PG commuting models is defined as (compare to our Eq. \ref{eq:Gdef}):
\begin{align}
\label{eq:twopoint}
 G_j(\tau) =  2Z^{-1}_\beta \,\bra{I}
 {\rm e}^{-\beta H_L} \psi^\ell_j(\tau)  \psi^\ell_j(0)
 \ket{I} \;;\quad \psi^\ell_j(\tau) = {\rm e}^{\tau H_L} \psi^\ell_j
 {\rm e}^{-\tau H_L}
\end{align}
where the Majorana index $j$ is not summed over.
In \cite{Gao:2023gta} a closed form expression for the ensemble average of
this correlator is derived
in the approximation that the ensemble average of the numerator is performed
separately from the ensemble average of the partition function $Z$ that appears
in the denominator. 
The closed form expression is:
\begin{align}
\label{eq:Gapprox}
 \overline{G_j(\tau)} = {\rm exp}\left(
 -{\cal J}^2\tau(\beta - \tau )\right)
\end{align}
The same approximation is used in \cite{Gao:2023gta} to compute the OTOCs considered in the next section.

\begin{figure}[htb]
\begin{center}
\includegraphics[width=0.45\textwidth]{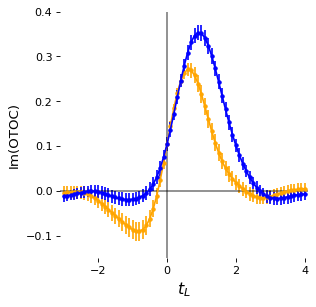}
\includegraphics[width=0.45\textwidth]{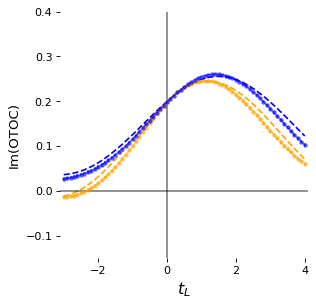}
\includegraphics[width=0.45\textwidth]{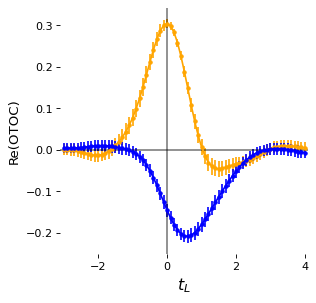}
\includegraphics[width=0.45\textwidth]{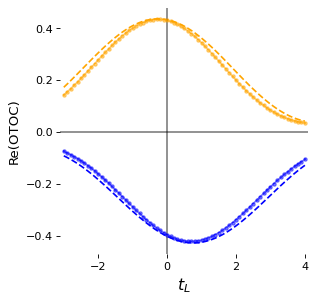}
\protect\caption{Comparison of $N$$=$$10$ PG commuting models (left) and
SYK models (right) using a direct ensemble averaging of 100 instantiations
of the finite $N$ OTOC,
taking $\beta = 1$, $\mu = \pm 0.139\pi$, and fixing $t_R = -0.720$.
Shown is $-$sgn$(\mu )\,H_{i\mu}(t_L,t_R)$ as a function of $t_L$,
for positive $\mu$ (blue points) and negative $\mu$ (orange points).
The upper plots show the imaginary parts, and the lower plots the real parts.
Also shown in the right plots are results (dashed lines)
for the Learned17 SYK-based model
introduced in \cite{jafferis2022traversable}.
\label{fig:compareSYKbeta10}}
\end{center}
\vspace{-0.5cm}
\end{figure}

\subsection{OTOCs}

The fast scrambling behavior of the SYK model can be exhibited in a variety
of out-of-time-order 4-point correlators (OTOCs). The analysis
in \cite{Gao:2023gta} (see eqn 4.1 in that paper) uses the following OTOC
in an attempt to connect to the properties of wormhole teleportation:
\begin{align}
\label{eq:OTOC-C}
 C(t_L,t_R) \equiv \langle {\rm tfd} | \{
 {\rm e}^{i\mu V} \psi^\ell_j(t_L) {\rm e}^{-i\mu V}, \psi^r_j(t_R) \}
 | {\rm tfd} \rangle
 = -2{\rm Im}(H_{i\mu}(t_L,t_R) )
\end{align}
where
\begin{align}
\label{eq:OTOC-H}
 H_{i\mu}(t_L,t_R) \equiv -i\,\langle {\rm tfd} | 
 {\rm e}^{i\mu V} \psi^\ell_j(t_L) {\rm e}^{-i\mu V} \psi^r_j(t_R)
 | {\rm tfd} \rangle
\end{align}
and the Majorana index $j$ here and elsewhere is not summed over; in the figures we use $j$$=$$2$.
As in \cite{Gao:2023gta} it is useful to also consider the Euclidean version:
\begin{align}
\label{eq:OTOC-h}
 h_{\mu}(\tau_1,\tau_2) \equiv -i\,\langle {\rm tfd} | 
 {\rm e}^{\mu V} \psi^\ell_j(\tau_1) {\rm e}^{-\mu V} \psi^r_j(\tau_2)
 | {\rm tfd} \rangle
\end{align}

Figure \ref{fig:compareSYKbeta10} shows our computation of
the finite $N$ OTOC $-$sgn$(\mu )\,$Im($H_{i\mu}(t_L,t_R)$),
Here $|\mu |$, $t_R$ and $\beta$ are fixed to 
values chosen in \cite{Gao:2023gta} to optimize the peaking behavior as a function
of $t_L$ and the asymmetry in the OTOC between negative and positive $\mu$ values, i.e., the parameters were tuned attempting to maximize the extent to which the non-holographic PG commuting models for a given $\beta$ can ``mimic'' one of the characteristic
behaviors of holographic wormhole teleportation.
Our result agrees very well with Figure 5(c) of \cite{Gao:2023gta}.
Since the computation of this OTOC in \cite{Gao:2023gta} uses the
same approximation as the computation of the 2-pt function (i.e. the partition
function is ensemble-averaged separately), we do not expect exact agreement.
We also used $N$$=$$10$ rather than $N$$=$$8$ as in \cite{Gao:2023gta}, which
can produce small differences.

For fixed $t_R$, the imaginary part of
the OTOC shows peaking as a function of $t_L$, and that there is 
a somewhat higher peak for positive values of $\mu$ versus negative values.
In \cite{Gao:2023gta} it was suggested that this behavior of the finite $N$ OTOC
implies that the PG commuting models to some extent mimic the behavior of SYK
wormhole teleportation, including the $\mu$ asymmetry that for wormhole
teleportation is connected to a negative energy pulse in the holographic dual
description (note that, for the conventions used in defining
this OTOC, positive $\mu$ would be expected to correspond
to a negative energy pulse in the large $N$ limit of Gao-Jafferis teleportation).
We will show in the next subsection that it is not the case for 
teleportation with the finite $N$ PG commuting models that we consider here.

Figure \ref{fig:compareSYKbeta10} shows a side-by-side comparison of $N$$=$$10$ PG commuting models with
SYK models, computing the same OTOC using a direct ensemble averaging of 100 instantiations. Here again we use the parameters chosen in 
\cite{Gao:2023gta} to maximize the extent to which
the non-holographic PG commuting models for a given $\beta$ can ``mimic'' one of the characteristic behaviors of wormhole teleportation.

We also show the OTOC computed for the Learned17 model
introduced in \cite{jafferis2022traversable}. 
This model appears similar to
the PG commuting models of \cite{Gao:2023gta}, since the 5 terms in each of the Left and Right side
Hamiltonians are mutually commuting. However the Majorana fermion groupings 
of the Learned17 model are such
that the Left and Right side Hamiltonians are not instantiations of the PG commuting models.
As already discussed in
\cite{Jafferis:2023moh}, there are a number of observations and consistency
checks indicating that this similarity is misleading, and that the Learned17 is SYK based. In Figure \ref{fig:compareSYKbeta10} we see additional
evidence of this: the OTOC for the Learned17 model tracks the 
behavior of the model using two copies of $N$$=$$10$ SYK, with no possibility of confusing
them with the corresponding ensemble averaged PG commuting model of
\cite{Gao:2023gta}. This is true in spite of the fact that we used
the same parameter choices as in \cite{Gao:2023gta}, which are far
away from those
used to obtain the Learned17 Hamiltonian in \cite{jafferis2022traversable}.

\begin{figure}[htb]
\vspace{-0.5cm}
\begin{center}
\includegraphics[width=0.55\textwidth]{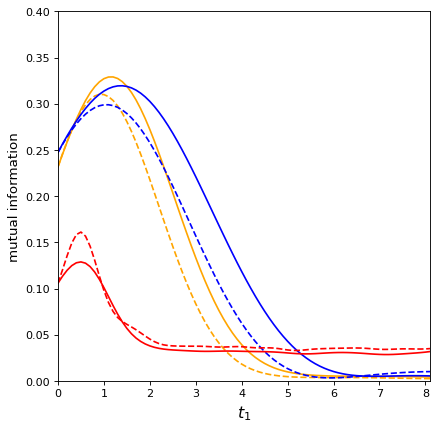}
\protect\caption{Comparison of the Gao-Jafferis
teleportation protocol using either two
copies of the $N$$=$$10$ SYK model, or two copies of the
$N$$=$$10$ PG commuting model.
Results are an ensemble average of 10 and 100 instantiations respectively, taking $\beta = 1$, $\mu = \pm 0.139\pi$, and fixing the
teleportation injection time as $t_0 = -0.720$.
Shown is the mutual information between the reference message qubit
$R$ and the extracted final state message qubit $T$ as a function of the
extraction time $t_1$. The PG commuting model results are shown for
negative $\mu$ (solid red) and positive $\mu$ (dashed red).
Results for the SYK model are shown for
negative $\mu$ (solid orange) and positive $\mu$ (dashed orange).
Also shown are results for the Learned17 SYK-based model
introduced in \cite{jafferis2022traversable}: for negative
$\mu$ (solid blue) and positive $\mu$ (dashed blue).
\label{fig:compareMIbeta10}} 
\end{center}
\end{figure}

\subsection{Teleportation}

We make a direct comparison of PG commuting and SYK-based models applying
the teleportation protocol of Gao and Jafferis. Such a 
comparison did not appear in \cite{Gao:2023gta}, where the analysis was
limited to calculations that can be performed analytically in a certain approximation for the ensemble averaging.
As for the OTOC comparisons of the previous section, we use the parameter
choices from \cite{Gao:2023gta}, which were tuned attempting to maximize the extent to which the non-holographic PG commuting models for a given $\beta$ can ``mimic'' behaviors of wormhole teleportation. Of course one does not
expect to optimize the wormhole teleportation features of 
the SYK-based models
for these same choices of parameters, so the comparison shown here
essentially maximizes the potential to confuse 
holographic and non-holographic dynamics in the same teleportation
protocol.

Figure \ref{fig:compareMIbeta10}
shows the comparison for $\beta = 1$, with the mutual information
plotted as a function of the extraction time $t_1$ for a fixed value
of the injection time $t_0$. The $N$$=$$10$ SYK model shows the expected
wormhole teleportation peak, with a larger peak for negative values of
$\mu$, consistent with a negative energy pulse in the holographic dual
description. The corresponding $N$$=$$10$ PG commuting model shows a
much smaller peak, at a significantly earlier time, and with different
late time behavior. Notably, the PG commuting model has a larger peak
for positive values of $\mu$ rather than negative values, inconsistent
with a holographic interpretation. In the same figure we also show
results for the Learned17 model.  As noted in the previous subsection, 
the Learned17 has a superficial similarity to
the PG commuting models of \cite{Gao:2023gta}. However, we see in Figure \ref{fig:compareMIbeta10} 
that the teleportation behavior of the learned model tracks that of the
SYK model, not that of the PG commuting model. We repeat that 
learned model
was originally derived from the $N$$=$$10$ SYK model 
with $\beta = 4$
and (in the conventions we use here) $\mu = \pm 0.095\pi$, with
$t_0 = -2.8$; nothing about the learning procedure, other than underlying holographic physics, 
would promote similarity to the SYK model for the parameter choices of Figure \ref{fig:compareMIbeta10}.

\begin{figure}[htb]
\vspace{-0.7 cm}
\begin{center}
\includegraphics[width=0.39\textwidth]{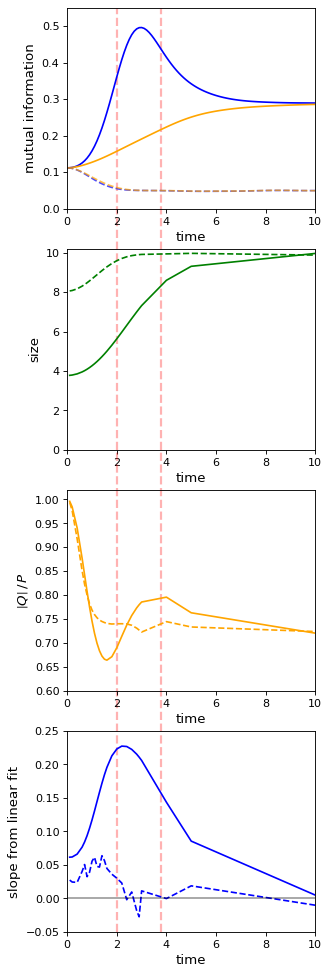}
\protect\caption{Comparison of PG commuting and SYK-based models. Results for the Gao-Jafferis protocol using a single instantiation 
of the SYK model
with $N$$=$$20$, $q$$=$$4$, ${\cal J}$$=$$1$, $\beta$$=$$4$, 
and using $\mu$$=\pm 0.2$,
compared with an ensemble
average of 100 instantiations of the PG commuting model of \cite{Gao:2023gta}
with the same parameter choices. Solid/dashed lines correspond to SYK/PG commuting models.
{\bf Top:} The mutual information
$I(R$$:$$T)$ as a function of time $t_0$, for negative/positive values of $\mu$
(blue/orange lines). 
{\bf Second:} The mean size of the
thermal fermion operator.
{\bf Third:} $|Q|/P$ averaged over size.
{\bf Bottom:} The fitted linear slope of the phases of $Q^\ell(s)$.
\label{fig:all-commuting}}
\end{center}
\end{figure}

\begin{figure}[htb]
\begin{center}
\includegraphics[width=0.4\textwidth]{figs/N20_size_winding_q4_Js1_beta4_mu020_sizes_t29.png}
\includegraphics[width=0.4\textwidth]
{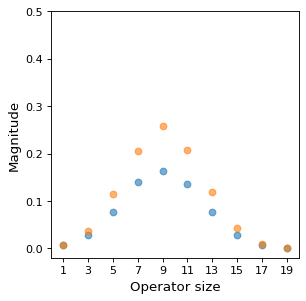}
\protect\caption{Comparison of the distribution of the thermal fermion operator
size. {\bf Left:} The Gao-Jafferis
protocol for a single instantiation of the SYK model with $N$$=$$20$,
$q$$=$$4$, $\beta$$=$$4$, ${\cal J}$$=$$1$, and taking $t_0$$=$2.9. 
Shown are 
$|Q^\ell(s)|$ (blue) and
$P^\ell(s)$ (orange) as a function of the operator size $s$; this is the same as
Figure \ref{fig:q4t29sizes} but showing only the first fermion.
{\bf Right:} Ensemble average of 100 instantiations of the
$N$$=$$20$ PG commuting model, also with
$q$$=$$4$, $\beta$$=$$4$, ${\cal J}$$=$$1$, and taking $t_0$$=$1.4.
The time chosen corresponds to the maximum slope of the phases from
a weighted linear fit.
\label{fig:q4t29sizes_commuting}}
\end{center}
\vspace{-0.4cm}
\end{figure}

\begin{figure}[htb]
\begin{center}
\includegraphics[width=0.33\textwidth]{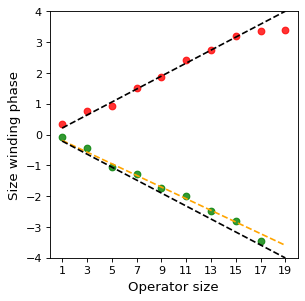}
\includegraphics[width=0.33\textwidth]{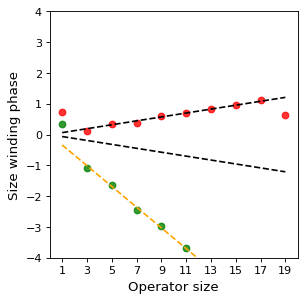}
\includegraphics[width=0.33\textwidth]{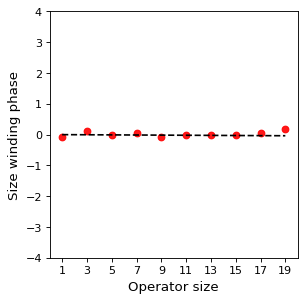}
\protect\caption{Size winding comparison for $N$$=$$20$ SYK-based and PG commuting models.
{\bf Upper left:} Size winding of the Gao-Jafferis
protocol for a single instantiation of the SYK model with $N$$=$$20$,
$q$$=$$4$, $\beta$$=$$4$, ${\cal J}$$=$$1$, $\mu$$=$$-0.2$, and taking $t_0$$=$2.9.
This is the same example as in Figure \ref{fig:q4t29comp} but showing only
the size winding of the first fermion.
{\bf Upper right:} Ensemble average of 100 instantiations of the
$N$$=$$20$ PG commuting model, also with
$q$$=$$4$, $\beta$$=$$4$, ${\cal J}$$=$$1$, and taking $t_0$$=$1.4.
The time chosen corresponds to the maximum slope of the phases from
a weighted linear fit (upper black dashed line).
The lower black dashed line is where the corresponding right size winding
should lie, while the orange dashed line is a linear fit to the actual
points after applying the Left-Right interaction.
{\bf Lower:} The same ensemble average of PG commuting models, looking now
at time $t_0$$=$2.9, and averaged over the first 10 fermions.
The black dashed line is a linear best fit with slope
-0.002  $\pm 0.006$, i.e., consistent with an
absence of size dependence in the phases of $Q^\ell(s)$.
\label{fig:commuting-sw}}
\end{center}
\vspace{-0.1cm}
\end{figure}

\subsection{Size winding}

It is interesting to check if the PG commuting models can mimic the
physics of size winding for parameter choices
in the same range where the SYK-based models exhibit the size winding
mechanism for quantum information transfer.
In Figure \ref{fig:all-commuting} we show an average over 100 
instantiations of the PG commuting model \cite{Gao:2023gta}, using
the same choice of parameters as for the Gao-Jafferis SYK-based example shown
in Figure \ref{fig:q4t29comp}, i.e.,
$N$$=$$20$, $q$$=$$4$, $\beta$$=$$4$, ${\cal J}$$=$$1$, $\mu$$=\pm 0.2$.
The top plot shows a comparison of the mutual information
$I(R$$:$$T)$ for the two models. The Gao-Jafferis model shows the usual
peaking behavior for negative $\mu$, while the PG commuting model shows
no peaking and no asymmetry at all. Both models show the same nonzero $I(R$$:$$T)$ at $t_0$$=$$0$,
corresponding to the direct transmission mechanism discussed in
section \ref{sec:warmup}. For the PG commuting model the contributions
to $I(R$$:$$T)$ appear to be a combination of the direct transmission
and ``peaked-size teleportation''. As is apparent in the second plot
of Figure \ref{fig:all-commuting} and in
Figure \ref{fig:q4t29sizes_commuting} (Right),
the PG commuting model has an earlier scrambling time, $t_0$$\simeq$$2$,
and already for $t_0$$=$$1.4$ the thermal fermion operator
size is peaked around $N$$/$$2$. At this same early time
the PG commuting model shows some linear size dependence
of the phases of $Q^\ell (s)$, as seen in
Figure \ref{fig:commuting-sw} (Upper right); this possibility
was already noted in \cite{Gao:2023gta}. However the slope
is much too small for the Left-Right interaction to convert
Left size winding to Right size winding, so size winding is not
a good description of the transfer of quantum information.

As already pointed out in section \ref{sec:warmup},
the direct transmission mechanism completely dies off by the scrambling
time; thus for $t_0$$>$$2$ we expect the PG commuting model
to display only ``peaked-size  teleportation''. The flat 
behavior of $I(R$$:$$T)$ of the PG commuting model for $t_0$$>$$2$ seen
in Figure \ref{fig:q4t29comp} (Top) is consistent with this.
We furthermore see in Figure \ref{fig:q4t29comp} (Bottom)
that in the same regime there is no statistically significant
dependence of the phases of $Q^\ell (s)$ on size.
Thus for example with $t_0$$=$$2.9$, where the complete size winding
mechanism works well for the SYK-based model
- see Figure \ref{fig:commuting-sw} (Upper left) - the phases in
the PG commuting model shows no sign of size dependence -
see Figure \ref{fig:commuting-sw} (Lower).

None of the behaviors of the PG commuting model that we have displayed
here resemble the SYK-based Learned17 model, which as shown
in  \cite{jafferis2022traversable} and \cite{Jafferis:2023moh} exhibits excellent size winding.

\subsection{Conclusions}
In this appendix we have compared the 
interesting finite $N$ PG commuting models
of \cite{Gao:2023gta} with the finite $N$ SYK and learned models of \cite{jafferis2022traversable}, including for the choices of parameters determined in
\cite{Gao:2023gta} as most likely to produce confusing similarities between the
properties of these models, similarities that might confuse our ability to 
separate holographic from non-holographic behaviors when implementing the wormhole teleportation protocol.

Our results show no such confusion. This is a strong check that the
results of \cite{jafferis2022traversable} for finite $N$, including
the learned model used in the Google Sycamore experiment,
are consistent with the expected properties of holographic wormhole dynamics.

\end{document}